\documentclass{amsart}
\usepackage{amsthm,amsfonts,amsmath,amscd,amssymb,latexsym,epsfig}
\newcommand{\gl}{{\mathfrak g \mathfrak l}}

\renewcommand{\u}{{\mathfrak u}}
\newcommand{\su}{{\mathfrak s  \mathfrak u}}

\newcommand{\cx}{{\mathbb C}}
\newcommand{\diag}{\operatorname{diag}}

\newcommand{\tr}{\operatorname{tr}}
\newcommand{\Res}{\operatorname{Res}}

\newcommand{\Lie}{\operatorname{Lie}}

\newcommand{\supp}{\operatorname{supp}}

\newcommand{\TP}{{\rm T\mathbb{P}^1}}

\newcommand{\Pic}{\operatorname{Pic }}

\numberwithin{equation}{section}

\newtheorem{theorem}{Theorem}[section]

\newtheorem{lemma}[theorem]{Lemma}

\newtheorem{corollary}[theorem]{Corollary}

\newtheorem{proposition}[theorem]{Proposition}

\theoremstyle{remark}

\newtheorem{remark}[theorem]{Remark}

\newtheorem{definition}[theorem]{Definition}

\newtheorem{ack}{Acknowledgment}

\newcommand{\C}{{\mathbb{C}}}

\newcommand{\oC}{{\mathbb{C}}}

\newcommand{\oN}{{\mathbb{N}}}

\newcommand{\oP}{{\mathbb{P}}}

\newcommand{\oR}{{\mathbb{R}}}

\newcommand{\oZ}{{\mathbb{Z}}}

\newcommand{\sA}{{\mathcal{A}}}   
\newcommand{\sB}{{\mathcal{B}}}
\newcommand{\sC}{{\mathcal{C}}}   

\newcommand{\sE}{{\mathcal{E}}}

\newcommand{\sG}{{\mathcal{G}}}   
\newcommand{\sH}{{\mathcal{H}}}

\newcommand{\sM}{{\mathcal{M}}}   
\newcommand{\sN}{{\mathcal{N}}}
\newcommand{\sO}{{\mathcal{O}}}

\newcommand{\sS}{{\mathcal{S}}}
\newcommand{\sT}{{\mathcal{T}}}

\newcommand{\sV}{{\mathcal{V}}}
\newcommand{\sW}{{\mathcal{W}}}

\newcommand{\sY}{{\mathcal{Y}}}

\newcommand{\fL}{{\mathfrak{l}}}

\newcommand{\fS}{{\mathfrak{s}}}

\newcommand{\fU}{{\mathfrak{u}}}

\begin{document}

\title{Monopoles and clusters}
\author{Roger Bielawski}
\address{School of Mathematics\\ University of Leeds\\Leeds LS2 9JT, UK}
\address{Mathematisches Institut, Universit\"at G\"ottingen, G\"ottingen 37073, Germany} 




\begin{abstract} We define and study certain hyperk\"ahler manifolds which capture the asymptotic behaviour of the $SU(2)$-monopole metric in regions where monopoles break down into monopoles of lower charges. The rate at  which these new metrics approximate the monopole metric is exponential, as for the Gibbons-Manton metric.
\end{abstract}

\maketitle

\section{Introduction} The moduli space $\sM_n$ of framed $SU(2)$-monopoles of charge $n$ on $\oR^3$ is a complete Riemannian manifold topological infinity of which corresponds to monopoles of charge $n$ breaking down into monopoles of lower charges. This asymptotic picture is given in Proposition (3.8) in \cite{AtHi} which we restate here:
\begin{proposition} Given an infinite sequence of points of $\sM_n$, there exists a subsequence $m_r$, a partition $n=\sum_{i=1}^s n_i$ with $n_i>0$, a sequence of points $x^i_r\in\oR^3$, $i=1,\dots,s$, such that
\begin{itemize}
\item[(i)] the sequence $m_r^i$ of monopoles translated by $-x_r^i$ converges weakly to a monopole of charge $n_i$ with centre at the origin;
\item[(ii)] as $r\rightarrow\infty$, the distances between any pair of points $x^i_r,x^j_r$ tend to $\infty$ and the direction of the line $x^i_rx^j_r$ converges to a fixed direction.
\end{itemize}
\label{first}\end{proposition} We can think of clusters of charge $n_i$ with centres at $x_r^i$ receding from one another in definite
directions.
\par
The aim of this paper is to capture this asymptotic picture in metric terms.   Observe that the above description, which leads to the
asymptotic metric being the product metric on $\prod \sM_{n_i}$, is valid only at infinity. It ignores the interaction of clusters at
finite distance from each other, e.g. the relative electric charges arising from their motion.  A physically meaningful description of the
asymptotic metric should take into consideration the contributions made by this interaction. Such an asymptotic metric, governing the
motion of dyons, was found by Gibbons and Manton \cite{GM} in the case when all $n_i$ are $1$, i.e. a monopole breaks down into particles.
It was then shown in \cite{CMP1} that this metric is an exponentially good approximation to the monopole metric in the corresponding
asymptotic region.
\par
Our aim is to generalise this to clusters of arbitrary charges. For any partition $n=\sum_{i=1}^s n_i$ with $n_i>0$ we define a space of
(framed) clusters $\sM_{n_1,\dots,n_s}$ with a natural (pseudo)-hyperk\"ahler metric. The picture is that as long as the size of clusters
is bounded, say by $K$ and the distances  between their centres $x^i$ are larger than some $R_0=R_0(K)$, then there are constants
$C=C(K),\alpha=\alpha(K)$ such that the cluster metric in this region of $\sM_{n_1,\dots,n_s}$ is $Ce^{-\alpha R}$-close to the monopole
metric in the corresponding region of $\sM_n$, where $R=\min \{|x^i-x^j|; \enskip i,j=1,\dots,s, i\neq j\}$.
\par
The definition of the cluster metric is given in terms of spectral curves and sections of the line bundle $L^2$, analogous to one of the definitions of the monopole metric (cf. \cite{AtHi}). Essentially, a framed cluster in  $\sM_{n_1,\dots,n_s}$ corresponds to $s$ real spectral curves $S_i$ of degrees $n_i$ together with  {\em meromorphic} sections of $L^2$ on each $S_i$, such that the zeros and poles of the sections occur only at the intersection points of different curves (together with certain nonsingularity conditions).
\par
Let us say at once that we deal here almost exclusively with the case of two clusters. Apart from notational complications when $s>2$, the
chief difficulty (also for $s=2$) is that unlike in the case of the Gibbons-Manton metric, we have not found a description of
$\sM_{n_1,\dots,n_s}$ as a moduli space of Nahm's equations. For $s=2$ we have such a description of the smooth (and complex) structure of
$\sM_{n_1,n_2}$ but not of its metric nor of the hypercomplex structure. The fact that our spaces of clusters $\sM_{n_1,\dots,n_s}$ are defined in terms
of spectral curves satisfying certain transcendental conditions, makes them quite hard to deal with. In particular, for $s>2$ we do
not have a proof that such curves exist (although we are certain that they do). For $s=2$ we do have existence, since the spectral curves in this case turn out to be spectral curves of $SU(2)$-calorons of charge $(n_1,n_2)$.  

\tableofcontents

\section{Line bundles and flows on spectral curves\label{lb}}

We recall here essential facts about spectral curves and line bundles. For a more detailed overview  we refer to \cite{reducible}.

\subsection{Line bundles and matricial polynomials}

In what follows $T$ denotes the total space of the line bundle $\sO(2)$ on $\oP^1$ ($T\simeq T\oP^1$), $\pi:T\rightarrow \oP^1$ is the
projection, $\zeta$ is the affine coordinate on $\oP^1$ and $\eta$ is the fibre coordinate on $T$. In other words $T$ is obtained by gluing
two copies of $\oC^2$ with coordinates $(\zeta,\eta)$ and $(\tilde{\zeta},\tilde{\eta})$ via:
$$ \tilde{\zeta}=\zeta^{-1}, \quad \tilde{\eta}=\eta/\zeta^2.$$
We denote the corresponding two open subsets of $T$ by $U_0$ and $U_\infty$.

Let $S$ be an algebraic curve in the linear system $\sO(2n)$, i.e. over $\zeta\neq \infty$ $S$ is defined by the equation
\begin{equation} P(\zeta,\eta)= \eta^n+a_1(\zeta)\eta^{n-1}+\cdots +a_{n-1}(\zeta)\eta+ a_n(\zeta)=0,\label{S}\end{equation}
where  $a_i(\zeta)$ is a polynomial of degree $2i$. $S$ can be singular or non-reduced (although spectral curves corresponding to monopoles, or to the clusters considered here are always reduced).
\par
We recall the following facts (see, e.g., \cite{Hi,AHH}):
\begin{proposition} The group $H^1(T,\sO_T)$ (i.e. line bundles on $T$ with zero first Chern class) is generated by  $\eta^i\zeta^{-j}$, $i>0$, $0<j<2i$. The corresponding line bundles have transition functions $\exp(\eta^i\zeta^{-j})$ from $U_0$ to $U_\infty$.\hfill $\Box$\label{T}\end{proposition}
\begin{proposition} The natural map $H^1(T,\sO_T)\rightarrow H^1(S,\sO_S)$ is a surjection, i.e. $H^1(S,\sO_S)$ is generated by $\eta^i\zeta^{-j}$, $0<i\leq n-1$, $0<j<2i$.\hfill $\Box$\label{all}\end{proposition}

Thus, the (arithmetic) genus of $S$ is $g=(n-1)^2$. For a smooth $S$, the last proposition describes line bundles of degree $0$ on $S$.
In general, by a line bundle we mean an invertible sheaf and by a divisor we mean a Cartier divisor. The degree of a line bundle is defined  as its Euler characteristic plus $g-1$. The theta divisor $\Theta$ is  the set of line bundles  of degree $g-1$ which have a non-zero section.
\par
Let $\sO_T(i)$ denote the pull-back of $\sO(i)$ to $T$ via $\pi:T\rightarrow \oP^1$. If $E$ is a sheaf on $T$ we denote by $E(i)$ the sheaf
$E\otimes \sO_T(i)$ and similarly for sheaves on $S$. In particular, $\pi^\ast \sO$ is identified with $\sO_S$.
\par
If $F$ is a line bundle of degree $0$ on $S$, determined by a cocycle $q\in H^1(T,\sO_T)$, and $s\in H^0\bigl(S, F(i)\bigr)$, then we denote by
$s_0,s_\infty$ the representation of $s$ in the trivialisation $U_0,U_\infty$, i.e.:
\begin{equation} s_\infty(\zeta,\eta)=\frac{e^q}{\zeta^i}s_0(\zeta, \eta).\label{represent}\end{equation}

 We recall the following theorem of Beauville \cite{Beau}:
\begin{theorem} There is a $1-1$ correspondence between the affine Jacobian $J^{g-1}-\Theta$ of line bundles of degree $g-1$ on $S$ and $GL(n,\cx)$-conjugacy classes of $\gl(n,\cx)$-valued polynomials $A(\zeta)=A_0+A_1\zeta+A_2\zeta^2$ such that $A(\zeta)$ is regular for every $\zeta$ and the characteristic polynomial of $A(\zeta)$ is \eqref{S}.\hfill ${\Box}$\label{Beauville} \end{theorem}

The correspondence is given by associating to a line bundle $E$ on $S$ its direct image $V=\pi_\ast E$, which has a structure of a $\pi_\ast \sO$-module. This is the same as a homomorphism $A:V\rightarrow V(2)$ which satisfies \eqref{S}. The condition $E\in J^{g-1}-\Theta$ is equivalent to $H^0(S,E)=H^1(S,E)=0$ and, hence, to $H^0(\oP^1,V)=H^1(\oP^1,V)=0$, i.e. $V=\bigoplus \sO(-1)$. Thus, we can interpret $A$ as a matricial polynomial precisely when $E\in J^{g-1}-\Theta$.

Somewhat more explicitly, the correspondence is seen from the exact sequence
\begin{equation} 0\rightarrow \sO_T(-2)^{\oplus n}\rightarrow \sO_T^{\oplus n}\rightarrow E(1)\rightarrow 0, \label{bundle}\end{equation}
where the first map is given by $\eta\cdot 1-A(\zeta)$ and $E(1)$ is viewed as a sheaf on $T$ supported on $S$. The inverse map is defined by the commuting diagram
\begin{equation}\begin{CD} H^0\bigl(S,E(1)\bigr) @>>> H^0\bigl(D_{\zeta}, E(1)\bigr)\\ @V \tilde{A}(\zeta) VV  @VV \cdot \eta V \\ H^0\bigl(S,E(1)\bigr) @>>> H^0\bigl(D_{\zeta}, E(1)\bigr), \end{CD}
\label{endom}\end{equation} where $D_{\zeta}$ is the divisor consisting of points of $S$ which lie above $\zeta$ (counting multiplicities).
That the endomorphism $\tilde{A}(\zeta)$ is quadratic in $\zeta$ is proved e.g. in \cite{AHH}. Observe that if $D_{\zeta_0}$ consists of
$n$ distinct points $p_1,\dots,p_n$ and if $\psi^1,\dots \psi^n$ is a basis of $H^0\bigl(S,E(1)\bigr)$, then $\tilde{A}(\zeta_0)$ in this basis is
\begin{equation} A(\zeta_0)=\left[ \psi^j(p_i)\right]^{-1} \diag\bigl(\eta(p_1),\dots,\eta(p_n)\bigr)\left[ \psi^j(p_i)\right],\label{conjugate} \end{equation}
where $\left[ \psi^j(p_i)\right]$ is a matrix with rows labelled by $i$ and columns by $j$.
\begin{remark} For a singular curve $S$, Beauville's correspondence most likely extends to $\overline{J^{g-1}}-\overline{\Theta}$, where $\overline{J^{g-1}}$ is the compactified Jacobian  in the sense of \cite{Alex}. It seems to us that this is essentially proved in \cite{AHH}.\label{comp}\end{remark}

Let $K$ be the canonical (or dualising) sheaf of $S$. We have $K\simeq  \sO_S(2n-4)$. If $E$ belongs to $J^{g-1} -\Theta$, then so does $E^\ast \otimes K$ and:
\begin{proposition} Let $A(\zeta)$ be the quadratic matricial polynomial corresponding to $E\in J^{g-1}-\Theta$. Then $A(\zeta)^T$ corresponds to $E^\ast \otimes K$. In particular, theta-characteristics outside $\Theta$ correspond to symmetric matricial polynomials.\label{canonical}\end{proposition}
For a proof, see \cite{reducible}.

\subsection{Real structure\label{realsect}} The space $T$ is equipped with a real structure (i.e. an antiholomorphic involution) $\tau$ defined by
\begin{equation} \zeta\mapsto -\frac{1}{\bar{\zeta}},\quad \eta\mapsto -\frac{\bar{\eta}}{\bar{\zeta}^2}. \label{antipodal}\end{equation}
Suppose that $S$ is real, i.e. invariant under $\tau$. Then $\tau$ induces an antiholomorphic involution $\sigma$ on $\Pic S$ as follows. Let $E$ be a line bundle on $S$ trivialised in a cover $\{U_\alpha\}_{\alpha\in A}$ with transition functions $g_{\alpha\beta}(\zeta,\eta)$ from $U_\alpha$ to $U_\beta$. Then $\sigma(E)$ is trivialised in the cover $\bigl\{\tau(U_\alpha)\bigr\}_{\alpha\in A}$ with transition functions
$$ \overline{g_{\alpha\beta}\bigl(\tau(\zeta,\eta)\bigr)}, $$
from $\tau(U_\alpha)$ to $\tau(U_\beta)$.  Observe that $\sigma(E)= \overline{\tau^\ast E}$
where ``bar" means taking the opposite complex structure. This map does not change the degree of $E$ and preserves line bundles $ \sO_S(i)$.
As there is a corresponding map on sections
\begin{equation} \sigma: s\mapsto \overline{\tau^\ast s},\label{sigma0}\end{equation}
it is clear that $J^{g-1} -\Theta$ is invariant under this map. The $\sigma$-invariant line bundles are called {\em real}. Real line bundles of degree $0$ have \cite{reducible}
transition functions $\exp q(\zeta,\eta)$, where $q$ satisfies:
$$\overline{q(\tau(\zeta,\eta))}=q(\zeta,\eta).$$
On the other hand, a line bundle $E$ of degree $d=in$, $i\in \oZ$, on $S$ is real if and only if it is of the form $E=F(i)$, where $F$ is a real line bundle of degree $0$.

For bundles of degree $g-1$ we conclude (see \cite{reducible} for a proof):
\begin{proposition} There is a $1-1$ correspondence between $J_{\oR}^{g-1}-\Theta_\oR$ and conjugacy classes of matrix-valued polynomials $A(\zeta)$ as in Theorem \ref{Beauville} such that there exists a hermitian $h\in GL(n,\cx)$ with
\begin{equation} hA_0h^{-1}=-A_2^\ast,\quad hA_1h^{-1}=A_1^\ast,\quad hA_2h^{-1}=-A_0^\ast.\label{h}\end{equation}
\label{realbundles}\end{proposition}

\subsection{Hermitian metrics\label{hermitian}}

Let $S$ be a real curve.
\begin{definition} A line bundle of degree $g-1$ on $S$ is called {\em definite} if it is in $J_{\oR}^{g-1}-\Theta_\oR$ and the matrix $h$ in \eqref{h}
can be chosen to be positive-definite. The subset of definite line bundles is denoted by $J^{g-1}_+$. \label{pos}\end{definition}

We easily conclude that there is a 1-1 correspondence between $J_+^{g-1}$ and $U(n)$-conjugacy classes of matrix-valued polynomials
$A(\zeta)$ as in Theorem \ref{Beauville} which in addition satisfy
\begin{equation} A_2=-A_0^\ast,\quad A_1=A_1^\ast. \label{+} \end{equation}

Definite line bundles have also the following interpretation (cf. \cite{Hi}):
\par
For $E=F(n-2)\in J_\oR^{g-1}$ the real structure induces an antiholomorphic isomorphism
\begin{equation}\sigma: H^0\bigl(S, F(n-1)\bigr)\longrightarrow H^0\bigl(S, F^{\ast}(n-1)\bigr),
\label{sigma}\end{equation} via the map \eqref{sigma0}. Thus, for   $v,w\in H^0\bigl(S,
F(n-1)\bigr)$,  $v\sigma(w)$ is a section of $\sO_S(2n-2)$ and so it can be uniquely written \cite{Hi,AHH} as
\begin{equation} c_0\eta^{n-1}+ c_1(\zeta)\eta^{n-2}+\dots +c_{n}(\zeta), \label{c_0}
\end{equation}
where the degree of $c_i$ is $2i$. Following Hitchin \cite{Hi}, we define a hermitian form on $H^0\bigl(S, F(n-1)\bigr)$ by
\begin{equation} \langle v,w\rangle=c_0.\label{form}\end{equation}
The following fact can be deduced from \cite{Hi}:
\begin{proposition} A line bundle $E=F(k-2)\in J_{\oR}^{g-1}-\Theta_\oR$ is definite if and only if the above form on $H^0\bigl(S, F(k-1)\bigr)$ is definite.\hfill $\Box$
\end{proposition}

Let $s,s^\prime$ be two sections of $F(n-1)$ on $S$. The form $\langle s, s^\prime\rangle$ is given by computing the section
$Z=s\sigma(s^\prime)$ of $\sO(2n-2)$ on $S$. Writing
$$ Z(\zeta,\eta)=c_0\eta^{n-1}+ c_1(\zeta)\eta^{n-2}+\dots +c_{n}(\zeta)$$
on $S$, we have $\langle s, s^\prime\rangle=c_0$. If $P(\zeta,\eta)=0$ is the equation defining $S$, then for any $\zeta_0$, such that
$S\cap \pi^{-1}(\zeta_0)$ consists of distinct points, we have
$$ c_0= \sum_{(\zeta_0,\eta)\in S} \Res\frac{Z(\zeta_0,\eta)}{P(\zeta_0,\eta)}.$$
Thus, if we write $(\zeta_0,\eta_1),\dots, (\zeta_0,\eta_n)$ for the points of $S$ lying over $\zeta_0$, then we have
\begin{equation} \langle s, s^\prime\rangle=\sum_{i=1}^n\frac{s(\zeta_0,\eta_i)\cdot\sigma(s^\prime)(\zeta_0,\eta_i)}{\prod_{j\neq i}
\bigl(\eta_i-\eta_j\bigr)}.\label{c0}\end{equation}
Therefore, one can compute $\langle s, s^\prime\rangle$ from the values of the sections at two fibres of $S$ over two antipodal points of
$\oP^1$ (as long as the fibres do not have multiple points).

\subsection{Flows\label{flows}} If we fix a tangent direction on $J^{g-1}(S)$, i.e. an element $q$ of $H^1(S,\sO_S)$, then the linear flow of line bundles on
$J^{g-1}(S)$ corresponds to a flow of matricial polynomials (modulo the action of $GL(n,\cx)$). We shall be interested only in the flow
corresponding to $[\eta/\zeta] \in H^1(S,\sO_S)$. Following the tradition, we denote by $L^t$ the line bundle on $T$ with transition
function $\exp(-t\eta/\zeta)$ from $U_0$ to $U_\infty$.
\par
For any line bundle $F$ of degree $0$ on $S$ we denote by $F_t$ the line bundle $F\otimes L^t$. We consider the flow $F_t(k-2)$ on
$J^{g-1}(S)$. Even if $F=F_0$ is in the theta divisor, this flow transports one immediately outside $\Theta$, and so we obtain a flow of
endomorphisms of $V_t=H^0\bigl(S,F_t(k-1)\bigr)$. These vector spaces have dimension $k$ as long as $F_t(k-2)\not\in \Theta$. We obtain an
endomorphism $\tilde{A}(\zeta)$ of $V_t$ as equal to multiplication by $\eta$ on $H^0(S\cap \pi^{-1}(\zeta), F_t(k-1)\bigr)$, where
$\pi:T\rightarrow \oP^1$ is the projection.
\par
To obtain a flow of matricial polynomials one has to trivialise the vector bundle $V$ over $\oR$ (the fibre of which at $t$ is $V_t$). This is a
matter of choosing a connection. If we choose the connection $\nabla^0$ defined by evaluating sections at $\zeta=0$ (in the trivialisation
$U_0,U_\infty$), then the corresponding matricial polynomial $A(t,\zeta)=A_0(t)+A_1(t)\zeta+A_2(t)\zeta^2$ satisfies \cite{Hi,AHH}
$$ \frac{d}{dt}A(t,\zeta)=\left[A(t,\zeta), A_2(t)\zeta\right].$$
As mentioned above, if $F$ is a real bundle, then $V$ has a natural hermitian metric \eqref{form} (possibly indefinite). The above
connection is not metric, i.e. it does not preserve the form  \eqref{form}. Hitchin \cite{Hi} has shown that the connection
$\nabla=\nabla^0+\frac{1}{2}A_1(t)dt$ is metric and that, in a $\nabla$-parallel basis, the resulting $A(t,\zeta)$ satisfies
$$\frac{d}{dt}A(t,\zeta)=\left[A(t,\zeta), A_1(t)/2+A_2(t)\zeta\right].$$
If the bundle $F(k-1)$ is positive-definite, then so are all $F_t(k-1)$. If the basis of sections is, in addition, unitary, then the
polynomials $A(t,\zeta)$ satisfy the reality condition \eqref{+}. If we write $A_0(t)=T_2(t)+iT_3(t)$ and $A_1(t)=2iT_1(t)$ for
skew-hermitian $T_i(t)$, then these matrices satisfy the Nahm equations:
\begin{equation}\dot{T}_i+\frac{1}{2}\sum_{j,k=1,2,3}\epsilon_{ijk}[T_j,T_k]=0\;,\;\;\;\;
i=1,2,3.\label{Nahm0}\end{equation}

\section{The monopole moduli space\label{monopole}}

The moduli space of $SU(2)$-monopoles of charge $n$ has a well-known description as a moduli space of solutions to Nahm's equations
\cite{Nahm,Hi}. From the point of view of section \ref{flows} monopoles correspond to spectral curves on which the flow $L^t(n-1)$ is
periodic and does not meet the theta divisor except for the periods. We can then describe the moduli space of $SU(2)$-monopoles as the
space of solutions to Nahm's equations \eqref{Nahm0} on $(0,2)$ with symmetry $T_i(2-t)=T_i(t)^T$ (cf. Proposition \ref{canonical}) and
satisfying appropriate boundary conditions.
\par
If we wish to consider the moduli space $\sM_n$ of {\em framed monopoles} (which is a circle bundle over the moduli space of monopoles) and
its natural hyperk\"ahler metric, then it is better to allow gauge freedom and introduce a fourth $\fU(n)$-valued function $T_0(t)$. Thus
we consider the following variant of  Nahm's equations:
\begin{equation}\dot{T}_i+[T_0,T_i]+\frac{1}{2}\sum_{j,k=1,2,3}\epsilon_{ijk}[T_j,T_k]=0\;,\;\;\;\;i=1,2,3.\label{Nahm}\end{equation}
The functions $T_0,T_1,T_2,T_3$ are $\fU(n)$-valued, defined on an interval and analytic. The space of solutions is acted upon by the gauge
group $\sG$ of $U(n)$-valued functions $g(t)$:
\begin{eqnarray} T_0&\mapsto & gT_0g^{-1}-\dot{g}g^{-1}\nonumber\\ T_i&\mapsto & gT_ig^{-1}\;,\;\;\qquad i=1,2,3.\label{action}\end{eqnarray}
To obtain $\sM_n$ we consider solutions analytic on $(0,2)$ which have simple poles at $0,2$, residues of which define a fixed irreducible
representation of $\su(2)$. The space $\sM_n$ is identified with the moduli space of solutions to \eqref{Nahm} satisfying these boundary
conditions and the symmetry condition $T_i(2-t)=T_i(t)^T$, $i=0,1,2,3$, modulo the action of gauge transformations $g(t)$ which satisfy
$g(0)=g(1)=1$ and $g(2-t)^{-1}=g^T(t)$.
\par
The tangent space at a solution $(T_0,T_1,T_2,T_3)$ can be identified with the space of solutions to the following system of linear
equations:
\begin{equation}\begin{array}{c} \dot{t}_0+[T_0,t_0]+[T_1,t_1]+[T_2,t_2]+[T_3,t_3]=0,\\
\dot{t}_1+[T_0,t_1]-[T_1,t_0]+[T_2,t_3]-[T_3,t_2]=0,\\
\dot{t}_2+[T_0,t_2]-[T_1,t_3]-[T_2,t_0]+[T_3,t_1]=0,\\
\dot{t}_3+[T_0,t_3]+[T_1,t_2]-[T_2,t_1]-[T_3,t_0]=0.\end{array}\label{tangent}\end{equation} The first equation is the condition that
$(t_0,t_1,t_2,t_3)$ is orthogonal to the infinitesimal gauge transformations and the remaining three are linearisations of \eqref{Nahm}.
Again, the symmetry condition $t_i(2-t)=t_i(t)^T$ holds.
\par
$\sM_n$ carries a hyperk\"ahler metric defined by
\begin{equation}\|(t_0,t_1,t_2,t_3)\|^2=-\sum_{i=0}^3\int_{0}^2\tr t_i^2(s) ds \label{metric},\end{equation}

\medskip

We now describe $\sM_n$ and its metric in terms of spectral curves. $\sM_n$ consists of pairs $(S,\nu)$ where $S\in |\sO(2n)|$  satisfies
\begin{equation} H^0\bigl(S, L^s(n-1)\bigr)=0 \quad\text{for $s\in (0,2)$},\label{generic0}\end{equation}
\begin{equation} L^2_{|S}\simeq \sO\label{triv}\end{equation}
and $\nu$ is a section of $L^2$ of norm $1$ (the norm is defined by   $\|\nu\|^2=\nu\sigma(\nu)\in \sO\simeq \cx$, where $\sigma$ is
defined as in \eqref{sigma0} without the sign). This last condition guarantees in particular that $L^s(n-1)\in J_+^{g-1}$ for $s\in
[0,2]$.
\begin{remark} In \cite{Hi} there is one more condition: that $S$ has no
multiple components. This, however, follows from the other assumptions. Namely, an $S$, satisfying all other conditions, produces a solution to Nahm's equations with boundary conditions of $\sM_n$. Thus, $S$ is a spectral curve of a monopole and cannot have multiple components.\label{multiple}\end{remark}
With respect to any complex structure, $\sM_n$ is biholomorphic to $\text{Rat}_n\bigl(\oP^1\bigr)$ - the space of based (mapping $\infty$
to $0$) rational maps of degree $n$ on $\oP^1$. If we represent an $(S,\nu)\in\sM_n$ in the patch $\zeta\neq \infty$ by a polynomial
$P(\eta,\zeta)$ and a holomorphic function $\nu_0(\eta,\zeta)$, then, for a given $\zeta_0$, the denominator of the corresponding rational
map is  $P(\eta,\zeta_0)$. The numerator can be identified \cite{Hu0}, when the denominator has distinct zeros, with the unique polynomial
of degree $n-1$ taking values $\nu_0(\eta_i,\zeta_0)$ at the zeros $\eta_i$ of the denominator.
\par
The complex symplectic form (i.e. $\omega_2+i\omega_3$ for $\zeta=0$) arising from the hyperk\"ahler structure is the standard form on
$\text{Rat}_n\bigl(\oP^1\bigr)$:
\begin{equation} \sum_{i=1}^n \frac{dp(\eta_i)}{p(\eta_i)}\wedge d\eta_i,\label{sympl_form}\end{equation}
where $p(z)/q(z)\in\text{Rat}_n\bigl(\oP^1\bigr)$ has distinct roots $\eta_i$.
\par
The K\"ahler form $\langle I_{\zeta_0}\cdot,\cdot\rangle$ where $I_{\zeta_0}$ is the complex structure corresponding to $\zeta_0\in \oP^1$
is given by the linear term in the expansion of \eqref{sympl_form} as power series in $\zeta-\zeta_0$.
\par
To complete the circle of ideas we recall, after Donaldson \cite{Don} and Hurtubise \cite{Hu0,Hu}, how to read off the section of $L^2$
from a solution to Nahm's equations.  The Nahm's equations \eqref{Nahm} can be written in the Lax pair
$\frac{d}{dt}A(t,\zeta)=\left[A(t,\zeta), A_\#(t,\zeta)\right]$, where $\zeta$ is an affine coordinate on $\oP^1$ and
$$ A(t,\zeta)= \bigl(T_2(t)+iT_3(t)\bigr)+2T_1(t)\zeta +\bigl(T_2(t)-iT_3(t)\bigr)\zeta^2,$$
$$A_\#(t,\zeta)=\bigl(T_0(t)+iT_1(t)\bigr)+\bigl(T_2(t)-iT_3(t)\bigr)\zeta.$$
In the case of monopoles, the residues at $t=0,2$ of $A(t)$ and of $A_\#(t)$ define  irreducible representations of $\fS\fL(2,\oC)$, which
are independent of the solution. In addition, the $-(n-1)/2$-eigenspace of the residue of $A_\#$ is independent of $\zeta$ and can be
chosen to be generated by the first vector of Euclidean basis of $\oC^n$. There is a unique solution $w(t,\zeta)$ of $\frac{d}{dt}w+A_\#
w=0$ satisfying $t^{-(n-1)/2}w(t,\zeta)\rightarrow (1,0,\dots,0)^T$ as $t\rightarrow 0$. The rational map, for any $\zeta\neq\infty$,
corresponding to a solution to Nahm's equations is then $w(1,\zeta)^T \bigl(z-A(1,\zeta)\bigr)^{-1}w(1,\zeta)$. Thus the section of $L^2$,
which is the numerator of the rational map, is (in the patch $\zeta\neq \infty$)
\begin{equation}\nu_0=w(1,\zeta)^T \bigl(z-A(1,\zeta)\bigr)_{\text{adj}}w(1,\zeta).\label{numerator}\end{equation}

\section{The moduli space of two clusters\label{asym_space}}

We consider the space $\Sigma_{k,l}$ of pairs $(S_1,S_2)$ of compact, real curves $S_1\in |\sO(2k)|$, $S_2\in |\sO(2l)|$ such that there
exists a $D\subset S_1\cap S_2$ satisfying
\begin{itemize}
\item[(i)] $D\cup \tau(D)= S_1\cap S_2$ (as divisors).
\item[(ii)] Over $S_1$: $L^2[D-\tau(D)]\simeq \sO$; over $S_2$: $L^2[\tau(D)-D]\simeq \sO$.
\item[(iii)] $H^0\bigl(S_1, L^s(k+l-2)[-\tau(D)]\bigr)=0$ and $H^0\bigl(S_2, L^s(k+l-2)[-D]\bigr)=0$ for $s\in (0,2)$. In addition the first (resp. second) cohomology group vanishes also for $s=0$ if $k\leq l$ (resp. $l\leq k$).
\item[(iv)] $L^s(k+l-2)[-\tau(D)]$ on $S_1$ and $L^s(k+l-2)[-D]$ on $S_2$ are positive-definite in the sense of Definition \ref{pos} for every real $s$.
\end{itemize}

We now define the space $\sM_{k,l}$ as the set of quadruples $(S_1,\nu_1,S_2,\nu_2)$ where $(S_1,S_2)\in \Sigma_{k,l}$, $\nu_1$ and $\nu_2$
are sections of norm $1$ of $L^2[D-\tau(D)]$ on $S_1$ and of $L^2[\tau(D)-D]$ on $S_2$, respectively. The norm of a section is defined as
in the previous section (after \eqref{triv}).
\par
We observe that $\sM_{k,l}$ is a $T^2$-bundle over $\Sigma_{k,l}$ (this corresponds to a framing of clusters).
\par
 The space $\sM_{k,l}$ is should be viewed as a ``moduli space" of two (framed) clusters, of cardinality $k$ and $l$.  We shall show that $\sM_{k,l}$ is equipped with a (pseudo)-hyperk\"ahler metric. In the asymptotic region of $\sM_{k,l}$ the metric is positive-definite and exponentially close to the exact monopole metric in the region of $\sM_{k+l}$ where monopoles of charge $k+l$ separate into clusters of cardinality $k$ and $l$.
\par
There is of course the problem whether curves satisfying conditions (i)-(iii) above exist and finding enough of them to correspond to all
pairs of far away clusters. Recall that $\text{Rat}_m\bigl(\oP^1\bigr)$ denotes the space of based ($\infty\rightarrow 0$) rational maps on
degree $m$. We are going to show
\begin{theorem}  Let $\zeta_0\in\oP^1-\{\infty\}$. There exists a diffeomorphism from $\text{Rat}_k\bigl(\oP^1\bigr)\times\text{Rat}_l\bigl(\oP^1\bigr)$ onto an open
dense subset $\sM_{k,l}^{\zeta_0}$ of $\sM_{k,l}$ with the following property.  For every $\left(\frac{p_1(z)}{q_1(z)},
\frac{p_2(z)}{q_2(z)}\right) \in \text{Rat}_k\bigl(\oP^1\bigr)\times\text{Rat}_l\bigl(\oP^1\bigr)$  there exists a unique element
$(S_1,\nu_1,S_2,\nu_2)$ of $\sM_{k,l}^{\zeta_0}$ such that the polynomials  $P_i(\zeta,\eta)$ defining the curves $S_i$, $i=1,2$, satisfy
$P_i(\zeta_0,\eta)=q_i(\eta)$ and the values of $\nu_i$ at points of $\pi^{-1}(\zeta_0)\cap S_i$ (in the canonical trivialisation of
section \ref{flows}) are the values of the numerators $p_i$ at the roots of $q_i$.\label{existence}\end{theorem}
A proof of this theorem will be given at the end of the next section.\newline We can describe $\sM_{k,l}$ (but not its metric) as a moduli
space $N_{k,l}$ of solutions to Nahm's equations:
\begin{itemize}
\item[(a)] The moduli space consists of  $\fU(k)$-valued solutions $T_i^-$ on $[-1,0)$ and of $\fU(l)$-valued solutions $T_i^+$ on $(0,1]$.
\item[(b)] If $k\geq l$, then  $T_i^+$, $i=0,1,2,3$, $T_0^-$ and the $k\times k$ upper-diagonal block of $T_i^-$, $i=1,2,3$, are analytic at $t=0$. The $(k-l)\times(k-l)$ lower-diagonal blocks of $T^-_i$ have simple poles with residues defining the standard $(k-l)$-dimensional irreducible representation of ${\frak su}(2)$. The off-diagonal blocks of $T_i^-$ are of the form $t^{(k-l-1)/2}\times(\text{\it analytic in $t$})$. Similarly, if $l\geq k$, then $T_i^-$, $i=0,1,2,3$, $T_0^+$ and the $l\times l$ upper-diagonal block of $T_i^+$, $i=1,2,3$, are analytic at $t=0$; The $(l-k)\times(l-k)$  lower-diagonal blocks of $T^+_i$ have simple poles with residues defining the standard $(l-k)$-dimensional irreducible representation of ${\frak su}(2)$ and the off-diagonal blocks of $T_i^+$ are of the form  $t^{(l-k-1)/2}\times(\text{\it analytic in $t$})$.
\item[(c)] We have the following matching conditions at $t=0$: if $k<l$ (resp. $k>l$) then the limit  of the $k\times k$ upper-diagonal block of $T_i^+$ (resp. $l\times l$ upper-diagonal block of $T_i^-$)  at $t=0$ is equal to the limit of $T_i^-$ (resp. $T_i^+$) for $i=1,2,3$.
If $k=l$, then there exists a vector $(V,W)\in {\Bbb C}^{2k}$ such that $(T_2^++iT_3^+)(0_+)-(T_2^-+iT_3^-)(0_-)= VW^T$ and
$T_1^+(0_+)-T_1^-(0_-)=(|V|^2-|W|^2)/2$.
\item[(d)] The solutions are symmetric at $t=-1$ and at $t=1$.
\item[(e)] The gauge group $\sG$ consists of gauge transformations $g(t)$ which are $U(k)$-valued on $[-1,0]$, $U(l)$-valued on $[0,1]$, are orthogonal at $t=\pm 1$ and satisfy the appropriate matching conditions at $t=0$: if $k\leq l$, then the upper-diagonal $k\times k$ block of $g(t)$ is continuous, the lower-diagonal block is identity at $t=0$ and the off-diagonal blocks vanish to order $(l-k-1)/2$ from the left. Similarly for $l\leq k$.
\end{itemize}

\begin{remark} It is known that $N_{k,l}$ is isomorphic to the moduli space of $SU(2)$-calorons, i.e. periodic instantons \cite{NyS, ChH}. The matching conditions at $t=0$ are those for $SU(3)$-monopoles (cf. \cite{HuMu}).\end{remark}

\begin{remark} If we omit the condition that the $T_i$ are symmetric at $\pm 1$ and allow only gauge transformations which are $1$ at $\pm 1$, then we obtain the space $F_{k,l}(-1,1)$ considered in \cite{CMP2}. Thus $N_{k,l}$ is the hyperk\"ahler quotient of $F_{k,l}(-1,1)$ by $O(k)\times O(l)$. \label{Fkl}
\end{remark}

We have
\begin{proposition} There is a natural bijection between $\sM_{k,l}$ and $N_{k,l}$.
\label{symmNahm}\end{proposition}
\begin{proof} According to \cite{HuMu} the flow $L^t(k+l-1)[-D]$ on $S_1$ and $S_2$ corresponds to a solution to Nahm's equations (with $T_0=0$) satisfying the matching conditions of $N_{k,l}$ at $t=0$. The condition (iii) in the definition of $\Sigma_{k,l}$ is equivalent to regularity of the solution on $(-2,0)$ and on $(0,2)$. Proposition \ref{canonical} implies that the condition that the $T_i$ are symmetric at $\pm 1$ corresponds to $L^{-1}(k+l-1)_{|S_1}[-D]$ and  $L^{1}(k+l-1)_{|S_2}[-D]$ being isomorphic to $P_1(k-1)$ and $P_2(l-1)$, where $P_1$ and $P_2$ are elements of order two in the real Jacobians of $S_1$ and $S_2$. Hence $L^{-1}(l)_{|S_1}[-D]\simeq P_1$ and $L^{1}(k)_{|S_2}[-D]\simeq P_2$. Squaring gives $L^{-2}(2l)\simeq [2D]$ on $S_1$ and $L^2(2k)\simeq [2D]$ on $S_2$. Using the relations $[D+\tau(D)]\simeq \sO(2l)$ on $S_1$ and $[ D+\tau(D)]\simeq \sO(2k)$ on $S_2$ shows the condition (d) in the definition of $N_{k,l}$ is equivalent to (ii) in the definition of $\Sigma_{k,l}$. Therefore there is a 1-1 correspondence between $\Sigma_{k,l}$ and the spectral curves arising from solutions to Nahm's equations in $N_{k,l}$. Now, a pair of spectral curves determines an element of $N_{k,l}$ only once we have chosen $\tau$-invariant isomorphisms $L^{-1}(l)_{|S_1}[-D]\simeq P_1$ and  $L^{1}(k)_{|S_2}[-D]\simeq P_2$ or, equivalently, isomorphisms in (ii) in the definition of $\Sigma_{k,l}$. Conversely, extending a solution to Nahm's equations, which belongs to $N_{k,l}$, by symmetry to $(-2,0)\cup (0,2)$ gives isomorphisms of (ii).
\end{proof}

The space $N_{k,l}$ carries a natural hyperk\"ahler metric, defined in the same way as for other moduli spaces of solutions to Nahm's
equations. This is not, however, the asymptotic monopole metric, which will be defined in section \ref{metricsection}.

\section{The complex structure of $N_{k,l}$\label{compl_N}}

As remarked above (Remark \ref{Fkl}), $N_{k,l}$ has a natural hyperk\"ahler structure. We wish to describe $N_{k,l}$ as a complex manifold
with respect to one of these complex structures (the $SO(3)$-action rotating $T_1,T_2,T_3$ guarantees that all complex structures are
equivalent). As usual, such a proof involves identifying the hyperk\"ahler quotient with the complex-symplectic quotient. We have not been
able to show that {\em all} complex gauge orbits are stable (or equivalently, given Remark \ref{Fkl}, that all $O(k,\cx)\times
O(l,\cx)$-orbits on $F_{k,l}(-1,1)$ are stable) and so we only describe an open dense subset of $N_{k,l}$.
\par
We set $\alpha=T_0+iT_1$ and $\beta=T_2+iT_3$. The Nahm equations can be then written as one complex and one real equation:
\begin{eqnarray} & &\frac{d\beta}{dt} = [\beta,\alpha]\label{complex}\\
 & &\frac{d\,}{dt}(\alpha+\alpha^\ast) =[\alpha^\ast,\alpha]+[\beta^\ast,\beta].\label{real}\end{eqnarray}
 We define $\sA_{k,l}$ as the space of solutions $(\alpha,\beta)=\bigl((\alpha^-,\alpha^+),(\beta^-,\beta^+)\bigr)$ to the complex equation \eqref{complex} on $[-1,0)\cup(0,1]$  satisfying condition (b) of the definition of  $N_{k,l}$. Moreover $\beta$ (but not necessarily $\alpha$) satisfies conditions (c) and (d) of that definition. The space $\sA_{k,l}$ is acted upon by the complexified gauge group $\sG^\cx$, i.e. the group of complex gauge transformations satisfying the matching conditions in part (e) of  the definition of  $N_{k,l}$. Denote by $N^\text{\rm r}_{k,l}$ and $\sA^\text{\rm r}_{k,l}$ the subsets where $\beta(\pm 1)$ are regular matrices. We have
 \begin{proposition}$N^\text{\rm r}_{k,l} =\sA^\text{\rm r}_{k,l}/\sG^\cx$. \label{stability}\end{proposition}
\begin{proof} Let $\sN_{k,l}$ be the space of solutions to \eqref{complex} and \eqref{real} satisfying the conditions (a)-(d) of the definition of $N_{k,l}$, so that $N_{k,l}=\sN_{k,l}/\sG$.
We have to show that in every $\sG^\cx$-orbit in $\sA_{k,l}$, there is a unique $\sG$-orbit of an element of $\sN_{k,l}$. First we rephrase
the problem. Denote by $\tilde{\sA}_{k,l}$ (resp. $\tilde{\sN}_{k,l}$) the set of solutions to \eqref{complex} (resp. to both
\eqref{complex} and \eqref{real}) on $(-2,0)\cup (0,2)$ satisfying the matching conditions of $\sA_{k,l}$ (resp. $\sN_{k,l}$) at $0$ and,
in addition, $\alpha^\pm(\pm 2-t)=\alpha^\pm(t)^T$, $\beta^\pm(\pm 2-t)=\beta^\pm(t)^T$. Denote by $\tilde{\sG}^\cx$ (resp.
$\tilde{\sG}$)  the group of complex (resp. unitary) gauge transformations which satisfy the matching conditions of $\sG^\cx$ (resp. $\sG$)
at $0$ and, in addition, $g(t)^{-1}=g(-2-t)^T$ if $t\leq 0$ and $g(t)^{-1}=g(2-t)^T$ if $t\geq 0$. We observe that
$$\tilde{\sA}_{k,l}/\tilde{\sG}^\cx=\sA_{k,l}/\sG^\cx\quad\text{and}\quad \tilde{\sN}_{k,l}/\tilde{\sG}=\sN_{k,l}/\sG.$$
Indeed, the maps from the left-hand to the right-hand spaces are simply restrictions to $[-1,0)\cup (0,1]$. To define the inverses, we can
use an element of $\sG^\cx$ or $\sG$ to make $\alpha_-(-1)$ and $\alpha_+(1)$ symmetric. We now extend the solutions to $(-2,0)\cup (0,2)$
by symmetry, i.e. we put $\bigl(\alpha_+(t),\beta_+(t)\bigr)=\bigl(\alpha_+(2-t)^T,\beta_+(2-t)^T\bigr)$ for $t\geq 1$ and similarly for
$(\alpha_-,\beta_-)$.
\par
We shall show that every $\tilde{\sG}^\cx$-orbit in $\tilde{\sA}^{\text{\rm r}}_{k,l}$ contains a unique $\tilde{\sG}$-orbit of an element
of $\tilde{\sN}_{k,l}$. We proceed along the lines of \cite{Hu}. Given an element of $\tilde{\sA}_{k,l}$ and an $h\in GL(m,\cx)/U(m)$,
where $m=\min (k,l)$, we can solve the real equation separately on $(-2,0)$ and on $(0,2)$ via  a (unique up to action of $\tilde{\sG}$)
pair of complex gauge trasformations $g_-$ on $[-2,0]$ and $g_+$ on $[0,2]$ such that
\begin{itemize}
\item[(i)] $g_-$  and $g_+$ satisfy the matching condition of $\sG^\cx$ at $t=0$; \item[(ii)]  the upper diagonal $m\times m$-blocks of $g_-(0)$  and of  $g_+(0)$ are both equal to $h$;
\item[(iii)] $g_-(-2)=g_-^T(0)^{-1}$ and $g_+(2)=g_+^T(0)^{-1}$.
\end{itemize}
This is shown exactly as in \cite{Don} and in \cite{Hu}. The condition (iii) and uniqueness guarantee that $g_-(t)^{-1}=g_-(-2-t)^T$  and
$g_+(t)^{-1}=g_+(2-t)^T$, so that $g_-$ and $g_+$ define an element of $\tilde{\sG}^\cx$. We now need to show that there is a unique $h\in
GL(m,\cx)/U(m)$ for which the resulting solutions to Nahm's equations will satisfy the matching conditions at $t=0$. i.e. that the jump
$\Delta(\tilde{\alpha}+\tilde{\alpha}^\ast)$ of the resulting $\tilde{\alpha}_\pm= g_\pm\alpha_\pm g_\pm - \dot{g}_\pm g^{-1}$ at $t=0$
will vanish.  To prove this we need to show two things: that the map $h\mapsto \tr
\bigl(\Delta(\tilde{\alpha}+\tilde{\alpha}^\ast)^2\bigr)$ is proper and that the differential of $h\mapsto
\Delta(\tilde{\alpha}+\tilde{\alpha}^\ast)$ is non-singular.
\par
To prove the properness of $h\mapsto \tr \bigl(\Delta(\tilde{\alpha}+\tilde{\alpha}^\ast)^2\bigr)$ we need Lemma 2.19 in \cite{Hu} in our
setting. We observe that Hurtubise's argument goes through as long as we can show that that logarithms of eigenvalues of $g_-(-1)^\ast
g_-(-1)$ and of $g_+(1)^\ast g_+(1)$ have a bound independent of $h$. The next two lemmas achieve this.
\begin{lemma} Let $B$ be a regular symmetric $n\times n$ matrix. The adjoint $O(n,\cx)$-orbit of $B$ is of the form $O(n,\cx)/\Gamma$ where $\Gamma$ is a finite subgroup of $O(n,\oR)$. \end{lemma}
\begin{proof} Since $B$ is regular, the stabiliser of $B$ in $GL(n,\cx)$ is the set of linear combinations of powers of $B$ and hence consists of symmetric matrices. Thus any $g$ which is orthogonal and stabilises $B$ satisfies $g^2=1$.
Decompose $g$ as $e^{ip}A$ where $p$ is real and skew-symmetric and $A$ real and orthogonal. Then $e^{ip}$ stabilises $ABA^{-1}$ and
repeating the argument we get $p=0$. Thus $\Gamma$ is a closed subgroup of  $O(n,\oR)$  consisting of elements, the square of which is $1$, hence
discrete, hence finite.
\end{proof}
\begin{lemma} Let $(\alpha_1,\beta_1)$ and $(\alpha_2,\beta_2)$ be two solutions to (real and complex) Nahm's equations on $[-a,a]$ which differ by a complex gauge transformation $g(t)$, i.e. $(\alpha_2,\beta_2)=g(\alpha_1,\beta_1)$. Suppose in addition that $g(0)$ is orthogonal and that $\beta_1(0)$ is a regular symmetric matrix. Then $1/M\leq \tr g^\ast(0)g(0)\leq M$, where $M\in [1,+\infty)$ depends only on $a$ and on the eigenvalues of $\beta_1(0)$.\label{bound}
\end{lemma}
\begin{proof} The previous lemma shows that, if $g(0)$ tends to infinity in $O(n,\cx)$, then so does $\beta_2(0)=g(0)\beta_1(0)g(0)^{-1}$ in $\gl(n,\cx)$. The proof of
Proposition 1.3 in \cite{BielAGAG} shows, however, that there is a constant $C=C(a)$ such that for any solution $(\alpha,\beta)$ to Nahm's
equations on $[-a,a]$, $\tr \beta^\ast(0)\beta(0)\leq C+\sum |d_i|^2$, where $d_i$ are the eigenvalues of $\beta(0)$.
\end{proof}
\par
It remains to prove that the differential of $h\mapsto \Delta(\tilde{\alpha}+\tilde{\alpha}^\ast)$ is non-singular. As in \cite{Hu}, we
choose a gauge in which $\alpha=\alpha^\ast$. Let $1+\epsilon \rho$ be an infinitesimal complex gauge transformation (i.e. $\rho\in \Lie
\tilde{\sG}^\cx$) preserving the Nahm equations with $\rho$ self-adjoint. The differential of $\Delta(\tilde{\alpha}+\tilde{\alpha}^\ast)$
is then $-2\Delta\dot{\rho}$. The fact that $\rho$ preserves the Nahm equations implies that $\rho$ satisfies, on both  $(-2,0)$ and
$(0,2)$, the equation
$$\ddot{\rho}=[\alpha^\ast,[\alpha,\rho]]+[\beta^\ast,[\beta,\rho]]-[[\beta^\ast,\beta],\rho].$$
We compute the $L^2$-norm of $(a,b)=(-\dot{\rho}+[\rho,\alpha], [\rho,\beta])$ on  an interval $[r,s]$ contained in either $[-2,0]$ or
$[0,2]$:
\begin{equation} \int_r^s\langle-\dot{\rho}+[\rho,\alpha],-\dot{\rho}+[\rho,\alpha]\rangle +\langle[\rho,\beta],  [\rho,\beta]\rangle=-\tr\dot{\rho}\rho\bigl|^{s}_r.\label{L2}\end{equation}
Since $\rho(\pm 1)$ is skew-symmetric and $\dot{\rho}(\pm 1)$ is symmetric, $\tr\dot{\rho}\rho$ vanishes at $\pm 1$. Were the jump of
$\dot{\rho}$ to vanish at $0$, we would get
$$\int_{-1}^0\bigl(\|a\|^2+\|b\|^2\bigr)+\int_{0}^1\bigl(\|a\|^2+\|b\|^2\bigr)=0,$$
and hence, in particular, $[\rho,\beta]=0$ on both $[-1,0]$ and on $[0,1]$. Then $\rho(1)$ commutes with $\beta(1)$. As $\beta(1)$ is a
regular symmetric matrix, its centraliser consists of symmetric matrices and hence $\rho(1)$ is both symmetric and skew-symmetric, hence
zero. For the same reason $\rho(-1)$ vanishes. We can now finish the proof as in \cite{Hu}. \end{proof}
\par
One can now identify $N_{k,l}^\text{\rm r}$ as a complex affine variety. It is not however a manifold and for our purposes it is sufficient
to identify a  subset of $N_{k,l}^\text{\rm r}$. We consider sets $\sA_{k,l}^\text{\rm rr}$ and the corresponding $N_{k,l}^\text{\rm rr}$
essentially consisting of those solutions $(\alpha,\beta)$ for which  $\beta_-(0)$ and $\beta_+(0)$ do not have a common eigenvector with a
common eigenvalue.  More precisely, if $k<l$ (resp. $k>l$) we require that there is no $(\lambda,v)\in\cx\times \cx^k$ (resp.
$(\lambda,v)\in\cx\times \cx^l$) such that $\beta_-(0)v=\lambda v$ (resp. $\beta_+(0)v=\lambda v$) and $\lim_{t\rightarrow 0}
(\beta_+(t)-\lambda)\tilde{v}=0)$ (resp. $\lim_{t\rightarrow 0} (\beta_-(t)-\lambda)\tilde{v}=0)$, where
$\tilde{v}=\left(\begin{array}{c}v\\0\end{array}\right)$. If $k=l$ and $\beta_+(0)-\beta_-(0)=VW^T$, we only require that $W^Tv\neq 0$ for
any eigenvector $v$ of $\beta_-(0)$ (if $V\neq 0$, this is equivalent to $\beta_-(0)$ and $\beta_+(0)$ not having  a common eigenvector
with a common eigenvalue).  We have:
\begin{proposition}  $N_{k,l}^\text{\rm rr}$  is biholomorphic to  $\text{Rat}_k\bigl(\oP^1\bigr)\times\text{Rat}_l\bigl(\oP^1\bigr)$.
\label{rat_pairs}\end{proposition}
\begin{proof} Given Proposition \ref{stability}, it is enough to show that  $\tilde{\sA}^{\text{\rm rr}}_{k,l}/\tilde{\sG}^\cx$ is biholomorphic to  $\text{Rat}_k\bigl(\oP^1\bigr)\times\text{Rat}_l\bigl(\oP^1\bigr)$.
\par
{\bf The case of $k<l$.} First of all, just as in \cite{Hu,CMP2}, we use a singular gauge transformation to make $\beta_+(0)$  regular and
of the form
\begin{equation}\beta_+(0)=\left(\begin{array}{ccc|cccc} &  &  & 0 &\ldots & 0 & g_1\\
& \beta_-(0) & & \vdots & & \vdots & \vdots\\ &  &  & 0 &\ldots & 0 & g_k\\ \hline
f_1 & \ldots & f_k & 0 &\ldots & 0 & e_1\\
0 &\ldots & 0 & 1 & \ddots & & e_2\\
\vdots & & \vdots & & \ddots & \ddots & \vdots \\0 &\ldots & 0 & 0&\ldots & 1 & e_{l-k} \end{array}\right).\label{beta+}\end{equation} The
quotient $\tilde{\sA}^{\text{\rm r}}_{k,l}/\tilde{\sG}^\cx$ becomes the quotient $\tilde{\sB}^{\text{\rm rr}}_{k,l}/\tilde{\sG}^\cx$ where
$\tilde{\sB}_{k,l}$ is defined exactly as $\tilde{\sA}_{k,l}$, except that the matching condition for $\beta$ at $t=0$ is now given by
\eqref{beta+}. The superscript {\em rr} means now that both $\beta_-(0)$ and $\beta_+(0)$ are regular and do not have a common eigenvector
with a common eigenvalue. Since $\beta_-(0)$ is a regular matrix, we can find an element of $\tilde{\sG}^\cx$ which conjugates it to the
form:
\begin{equation}\left(\begin{array}{cccc}  0 &\ldots & 0 & b_1\\
  1 & \ddots & & b_2\\
 & \ddots & \ddots & \vdots \\ 0&\ldots & 1 & b_{k} \end{array}\right).\label{beta-}\end{equation}
The remaining gauge freedom are gauge transformations in $\tilde{\sG}^\cx$ such that their upper-diagonal block $h$ at $t=0$ centralises
\eqref{beta-}. We want to use this gauge freedom to make $(f_1,\dots,f_k)$ equal to $(0,\dots,0,1)$.
\begin{lemma} Let $B$ be a matrix of the form \ref{beta-} and let $u=(u_1,\dots,u_k)$ be a covector. There exists an invertible matrix $X$ such that  $XBX^{-1}=B$ and $uX^{-1}=(0,\dots,0,1)$ if and only if $uv\neq 0$ for any eigenvector $v$ of $B$. If such an $X$ exists, then it is unique.\label{cent-conj}\end{lemma}
\begin{proof} Since $(0,\dots,0,1)$ is a cyclic covector for $B$, there exists a unique $X$ such that $[X,B]=0$ and $u=(0,\dots,0,1)X$. The problem is the invertibility of $X$. We can write $X$ as $\sum_{i=0}^{k-1} c_iB^i$ for some scalars $c_i$. If we put $B$ in the Jordan form, then it is clear that $\det X\neq 0$ if and only if  $\sum_{i=0}^{k-1} c_i\lambda^i\neq 0$ for any eigenvalue $\lambda$ of $B$. Let $v=(v_1,\dots,v_k)^T$ be an eigenvector for $B$ with the eigenvalue $\lambda$. We observe that $Bv=\lambda v$ and $v\neq 0$ implies that $v_k\neq 0$. Since $uv=(0,\dots,0,1)Xv=v_k\sum_{i=0}^{k-1} c_i\lambda^i$, we conclude that $\det X\neq 0$ precisely when $uv\neq 0$ for any eigenvector $v$.
\end{proof}

Returning to the proof of the proposition, we observe that the condition that $\beta_-(0)$ of the form \ref{beta-} and $\beta_+(0)$ of the
form \ref{beta+} do not have a common eigenvector with a common eigenvalue is equivalent to $(f_1,\dots,f_k)v\neq 0$ for any eigenvector
$v$ of $\beta_-(0)$. Thanks to the above lemma we can now find a unique gauge transformation in $\tilde{\sG}^\cx$ such that its upper-diagonal
block $h$ at $t=0$ centralises  \eqref{beta-} and
  which makes $(f_1,\dots,f_k)$ equal to $(0,\dots,0,1)$.  The only gauge transformations which preserve this form of $\beta_\pm(0)$ are those which are identity at $t=0$ (and hence at $t=\pm 2$). We can now find a unique pair $(g_-,g_+)$ of gauge transformations on $[-2,0]$ and $[0,2]$ with $g_\pm(0)=1$ which make $\alpha$ identically zero.
Therefore sending $(\alpha,\beta)$ to $\bigl(\beta_+(0),g_-(-2),g_+(2)\bigr)$ gives a well-defined map from $\tilde{\sA}^{\text{\rm
rr}}_{k,l}/\tilde{\sG}^\cx$ to the set of $(B_+,g_1,g_2)\in \gl(l,\cx)\times GL(k,\cx)\times GL(l,\cx)$, where $B_+$ is of the form
\eqref{beta+} with $\beta_-(0)$ of the form \eqref{beta-}, $(f_1,\dots,f_k)=(0,\dots,0,1)$, $g_1^{-1}\beta_-(0)g_1=\beta_-(0)^T$,
$g_2^{-1}B_+g_2=B_+^T$. Let us write $B_-$ for $\beta_-(0)$. We observe that giving $g_1$  with $g_1^{-1}B_-g_1=B_-^T$ is the same as
giving a cyclic covector $w_1$ for $B_-$. The corresponding $g_1$ is $\bigl((B_-^T)^{k-1}w_1^T,\dots, B_-^Tw_1^T, w_1^T\bigr)$. The pair
$(B_-,w_1)$ corresponds to an element of $\text{Rat}_k\bigl(\oP^1\bigr)$ via the map $(B_-,w_1)\mapsto w_1(z-B_-)^{-1}(1,\dots,0)^T$.
\par
We claim that $(B_+,g_2)$ also corresponds to a unique element of $\text{Rat}_l\bigl(\oP^1\bigr)$. This follows from
\begin{lemma} Let $B_+$ be a matrix of the form \eqref{beta+} with $\beta_-(0)$ of the form \eqref{beta-} and $(f_1,\dots,f_k)=(0,\dots,0,1)$. There exists an invertible matrix $A$, depending only on $\beta_-(0)$, which conjugates $B_+$ to an $l\times l$-matrix of the form \eqref{beta-}.\end{lemma}
\begin{proof} Since $B_+$ is regular we can represent it as multiplication by $z$ on $\cx[z]/\bigl(q_+(z)\bigr)$  where $q_+(z)=\det(z-B_+)$. Let $q_-(z)=\det(z-B_-)$. In the basis $1,z,\dots, z^{l-1}$, $B_+$ is of the form \eqref{beta-}, while in the basis $1,z,\dots, z^{k-1},q_-(z),zq_-(z),\dots, z^{l-k-1}q_-(z)$ it is of the form \eqref{beta+} with $\beta_-(0)$ of the form \eqref{beta-} and $(f_1,\dots,f_k)=(0,\dots,0,1)$.
\end{proof}
Therefore we can consider, instead of $(B_+,g_2)$, the pairs $\bigl(AB_+A^{-1}, Ag_2A^T\bigr)$ and proceed as for $(B_-,g_1)$.\newline {\bf
The case of $k>l$.} This is exactly symmetric to the previous case.\newline {\bf The case of $k=l$.} We have $\beta_+(0)-\beta_-(0)=VW^T$.
As in the case $k<l$ we conjugate $\beta_-(0)$ to the form \eqref{beta-}. By assumption $W^Tv \neq 0$ for any eigenvector $v$ of
$\beta_-(0)$, so Lemma \ref{cent-conj} shows that we can make $W^T$ equal to $(0,\dots,0,1)$ by a unique gauge transformation
$g(t)\in\tilde{\sG}^\cx$ such that $g(0)$ centralises $\beta_-(0)$.  It follows that $\beta_+(0)$ is also of the form \eqref{beta-}. The
remainder of the argument is basically the same (but simpler) as for $k<l$.
\end{proof}

We observe that the above proof identifies the complex symplectic form of $N_{k,l}^\text{\rm rr}$. If we ``double"the metric, i.e. consider
solutions on $(-2,0)\cup (0,2)$ (just as at the beginning of the proof of Proposition \ref{stability}) then the complex symplectic form is
given by
\begin{equation} \int_{-2}^0 \tr (d\alpha_-\wedge d\beta_-)+\int_{0}^2\tr (d\alpha_+\wedge d\beta_+)+\tr(dV\wedge dW^T),\label{symplectic}\end{equation}
where the last term occurs only if $k=l$. Since this form is invariant under {\em complex} gauge transformations, going trough the above
proof on the set where $\beta_-$ and $\beta_+$ have all eigenvalues distinct (compare also \cite{CMP1,CMP2}) shows that this form on
$\text{Rat}_k\bigl(\oP^1\bigr)\times\text{Rat}_l\bigl(\oP^1\bigr)$ is $-\omega_- +\omega_+$, where $\omega_\pm$ are standard forms on
$\text{Rat}_k\bigl(\oP^1\bigr)$ and $\text{Rat}_l\bigl(\oP^1\bigr)$, given on each factor by \eqref{sympl_form}.

\medskip

We can now prove the existence theorem \ref{existence}. For this we need to consider the correspondence in Proposition \ref{rat_pairs} for
different complex structures, i.e. for different $\zeta\in \oP^1$. This works essentially as in \cite{CMP1,CMP2} and shows that the
denominators of the rational maps trace curves $S_1,S_2$ in $\Sigma_{k,l}$ while the numerator of the first map gives a section $\kappa_1$ of
$L^{-2}[\tau(D)-D]$ and the numerator of the first map gives a section $\kappa_2$ of $L^{2}[\tau(D)-D]$. Setting $\nu_1=\sigma(\kappa_1)$
and $\nu_2=\kappa_2$ gives us an element of $\sM_{k,l}$. Since we had the correspondence between ({\em curves, sections}) and rational maps
for $N_{k,l}$, we have one for $\sM_{k,l}$.

\begin{remark} The proofs of \cite{CMP2} show that  a section of the twistor space of $N_{k,l}$ corresponding to $(S_1,\kappa_1,S_2,\kappa_2)$  will lie outside of $N_{k,l}^\text{\rm rr}$ for $\zeta\in \pi(S_1\cap S_2)$.\end{remark}

\section{The hyperk\"ahler structure of $\sM_{k,l}$\label{metricsection}}

The space $\sM_{k,l}$ has been defined in such a way that its hypercomplex structure is quite clear: the quadruples $(S_1,\nu_1,S_2,\nu_2)$
are canonically sections of a twistor space. We can describe this twistor space by changing the real structure (and, hence, sections) of the twistor space of $N_{k,l}$.
\par
 As already mentioned (Remark \ref{Fkl}), the  space $N_{k,l}$, being a moduli space of solutions to Nahm's equations has a natural
(singular) hyperk\"ahler structure. Let us double the metric on $N_{k,l}$ by considering solutions on $(-2,0)\cup (0,2)$ just as at the
beginning of the proof of Proposition \ref{stability}. Let $p:Z\bigl(N_{k,l}\bigr)\rightarrow \oP^1$ be the twistor space of this
hyperk\"ahler structure. The fibers of $p$ correspond to $N_{k,l}$ with different complex structures and so, by Proposition
\ref{rat_pairs}, each fiber has an open dense subset isomorphic to $\text{Rat}_k\bigl(\oP^1\bigr)\times\text{Rat}_l\bigl(\oP^1\bigr)$. The
real sections correspond to solutions of Nahm's equations and, by the arguments of the previous two sections, to quadruples
$(S_1,\kappa_1,S_2,\kappa_2)$, where $(S_1,S_2)\in \sS_{k,l}$, $\kappa_1$ is a norm $1$ section of $L^{-2}[\tau(D)-D]$ on $S_1$ and
$\kappa_2$ a norm $1$ section of $L^{2}[\tau(D)-D]$ on $S_2$ (at least on the open dense subset of $N_{k,l}$). Consider the mapping
\begin{equation} T:Z\bigl(N_{k,l}\bigr)\rightarrow Z\bigl(N_{k,l}\bigr)\end{equation}
defined in the following way. Let $\chi=(S_1,\kappa_1,S_2,\kappa_2)$ be the unique real section passing through a point $n\in
p^{-1}(\zeta)$ corresponding to the pair $(f_1,f_2)\in \text{Rat}_k\bigl(\oP^1\bigr)\times\text{Rat}_l\bigl(\oP^1\bigr)$. If
$\zeta\neq\infty$ and $\pi^{-1}(\zeta)\cap( S_1\cup S_2)$ consists of distinct points, then we can identify the numerator of $f_1$ with the
unique polynomial taking values $\kappa_1(\zeta,\eta_i)$ at points $\eta_i$ where $(\zeta,\eta_i)\in\pi^{-1}(\zeta)\cap S_1$ (where, once
again, we think of $\kappa_1$ as a pair of analytic functions in the standard trivialisation in $U_0,U_\infty)$). Define $T(n)\in
\pi^{-1}(\zeta)$ as $(g_1,g_2)\in \text{Rat}_k\bigl(\oP^1\bigr)\times\text{Rat}_l\bigl(\oP^1\bigr)$ where $g_2=f_2$, the denominator of
$g_1$ is the same as the denominator of $f_1$ and the numerator of $f_1$ is the unique polynomial taking values
$\sigma(\kappa_1)(\zeta,\eta_i)$ at points $\eta_i$ ($\sigma$ is given in \eqref{sigma}). We can extend $T$ by continuity to the remaining points of the fiber   $p^{-1}(\zeta)$
and, by doing the same over $U_\infty$, to $\zeta=\infty$. Observe that $T^2=\text{Id}$.
\par
Let $\tau$ denote the real structure of $Z\bigl(N_{k,l}\bigr)$. We define a new real structure by $\tau^\prime=T\circ \tau\circ T^{-1}$ and
define $Z$ as $Z\bigl(N_{k,l}\bigr)$ with real structure $\tau^\prime$. The points of $\sM_{k,l}$ are real sections of $Z$, since they are
of the form $T(\chi)$, where $\chi=(S_1,\kappa_1,S_2,\kappa_2)$ is a real section of $Z\bigl(N_{k,l}\bigr)$. The normal bundle of each
$T(\chi)$ must be direct sum of $\sO(1)$'s, since through every two points in distinct fibres there passes a unique section (as this is true for the normal bundle of $\chi$). Therefore we have a hypercomplex
structure on $\sM_{k,l}$. Finally, we modify the fibre-wise symplectic form on $Z\bigl(N_{k,l}\bigr)$ by taking $\omega_+ +\omega_-$ on
each fiber (compare with the remark after \eqref{symplectic}). This is an $\sO(2)$-valued symplectic form $\omega$ on $Z$ and evaluated on
real sections of $TZ$, $\omega$ gives real sections of $\sO(2)$. Thus we obtain a (pseudo)-hyperk\"ahler metric on $\sM_{k,l}$ (which may
be degenerate):
\begin{theorem} The space $\sM_{k,l}$ carries a canonical  hypercomplex structure. With respect to each complex structure an open dense subset of $\sM_{k,l}$
 can be identified with $\text{Rat}_k\bigl(\oP^1\bigr)\times\text{Rat}_l\bigl(\oP^1\bigr)$. In addition, there is a pseudo-hyperk\"ahler
 metric (with degeneracies) on $\sM_{k,l}$ compatible with the hypercomplex structure.
 The K\"ahler form corresponding to $\zeta_0$ of the hyperk\"ahler metric is given (on an open dense set, where the roots of each rational map are distinct) by the linear term in the power series expansion of
$$ \sum_{i=1}^k \frac{d\nu_1(\zeta,\eta_i)}{\nu_1(\zeta,\eta_i)}\wedge d\eta_i +\sum_{i=k+1}^{k+l}\frac{d\nu_2(\zeta,\eta_i)}{\nu_2(\zeta,\eta_i)}\wedge d\eta_i,$$
around $\zeta_0$, where $(\zeta,\eta_1),\dots,(\zeta,\eta_k)$ are the points of $\pi^{-1}(\zeta)\cap S_1$  and
$(\zeta,\eta_{k+1}),\dots,(\zeta,\eta_{k+l})$ are the points of $\pi^{-1}(\zeta)\cap S_2$.\label{twistor} \end{theorem}

\begin{remark} The above construction of a hypercomplex structure via a change of real structure of the twistor space can be seen already
in the twistor space description of Taub-NUT metrics in Besse \cite{Besse}, section 13.87. There a change of real structure leads to
replacing the Taub-NUT metric with a positive mass parameter to one with a negative mass parameter. It is know that the Taub-NUT metric with a
negative mass parameter {\em is} the asymptotic metric of charge $2$ monopoles \cite{AtHi,Man}.\label{TNUT}\end{remark}

\section{$\sM_{k,l}$ as a hyperk\"ahler quotient\label{hkquotient}}

We wish to expand Remark \ref{Fkl}.  
The moduli space $\sM_n$ of $SU(2)$-monopoles of charge $n$ can be obtained as a hyperk\"ahler quotient of a moduli space $\widehat{\sM}_n$ of $SU(n+1)$-monopoles with minimal symmetry breaking (see \cite{Dan} for the case $n=2$). Namely, $\widehat{\sM}_n$ is defined as the space of solutions to Nahm's equations on $(0,1]$, which have a simple pole at $t=0$ with residues defining the standard irreducible representation of $\su(2)$, modulo gauge transformations, which are identity at $t=0,1$. The gauge transformations which are orthogonal at $t=1$ induce an action of $O(n,\oR)$ on $\widehat{\sM}_n$ and $\sM_n$ is the hyperk\"ahler quotient of $\widehat{\sM}_n$ by $O(n,\oR)$. 
\par
  The nice thing about  $\widehat{\sM}_n$ is that the spectral curves involved do not need to satisfy any transcendental or even closed conditions: $\widehat{\sM}_n$ is a principal $U(n)$-bundle over an open subset of  all real spectral curves. We now define an analogous space for $\sM_{k,l}$. It should be viewed as given by generic pairs of spectral curves with framing being $U(k)\times U(l)$.
\par
We consider first the space $F_{k,l}$, already described in Remark \ref{Fkl}. It is defined in the same way as $N_{k,l}$ (cf. Section \ref{asym_space}), except that the condition (d) is removed and the orthogonality condition in (e) is replaced by $g(\pm 1)=1$. In other words, $F_{k,l}$ consists of $\u(k)$-valued solutions to Nahm's equations on $[-1,0)$ and of  $\u(l)$-valued solutions  on $(0,1]$, satisfying the matching conditions of $N_{k,l}$ at $t=0$, but arbitrary at $t=\pm 1$, modulo gauge transformations which are identity at $t=\pm 1$ (and satisfy the matching condition of $N_{k,l}$ at $t=0$. 
\par
$F_{k,l}$ is a hyperk\"ahler manifold \cite{CMP2} and $N_{k,l}$ is the hyperk\"ahler quotient of $F_{k,l}$ by $O(k,\oR)\times O(l,\oR)$ (the action is defined by allowing gauge transformations which are orthogonal at $t=\pm 1$).  The set of spectral curves, defined by elements of $F_{k,l}$, is given by:
\begin{definition} We denote by ${\sS}_{k,l}$ the space of  pairs $(S_1,S_2)$ of  real curves $S_1\in |\sO(2k)|$, $S_2\in |\sO(2l)|$, of the form \eqref{S}, without common components, such that  $S_1\cap S_2=D+\tau(D)$, $\supp D\cap \supp \tau(D)=\emptyset$, so that
\begin{itemize}
\item[(i)] $H^0\bigl(S_1, L^t(k+l-2)[-\tau(D)]\bigr)=0$ and $H^0\bigl(S_2, L^t(k+l-2)[-D]\bigr)=0$ for $t\in (0,1]$. In addition, if $k\leq l$ (resp. $l\leq
k$), then $H^0\bigl(S_1, \sO(k+l-2)[-\tau(D)]\bigr)=0$ (resp. $H^0\bigl(S_2, \sO(k+l-2)[-D]\bigr)=0$).
\item[(ii)] $L^t(k+l-2)[-\tau(D)]$ on $S_1$ and $L^t(k+l-2)[-D]$ on $S_2$ are positive-definite in the sense of Definition \ref{pos} for every
$t$.
\end{itemize}
\label{Skl}\end{definition}
One can show that $F_{k,l}$ is a $U(k)\times U(l)$-bundle over ${\sS}_{k,l}$, but we shall not need this. What we do need is the complex structure of $F_{k,l}$ or, rather, its open subset $F^{\rm rr}_{k,l}$, defined in exactly the same way as  $N^{\rm rr}_{k,l}$. As in Section \ref{compl_N}, we fix a complex structure and write Nahm's equations as the complex one and the real one. According to \cite{CMP2},  $F_{k,l}$ is biholomorphic to $W\times GL(l,\cx)$, where , for $k<l$, $W$ is the set of matrices of the form \eqref{beta+}, while for $k=l$, $W$ is the set $\{(B_-,B_+,V,W)\in \gl(l,\cx)^2\times (\cx^l)^2; B_+-B_-=VW^T\}$. Thus $F_{k,l}$ is biholomorphic to $GL(l,\cx)\times \gl(k,\cx)\times \cx^{k+l}$. On the other hand, the proof of Proposition \ref{rat_pairs} furnishes a different biholomorphism for $F^{\rm rr}_{k,l}$: 
\begin{proposition} $F^{\rm rr}_{k,l}$ is biholomorphic to $\cx^k\times GL(k,\cx)\times \cx^l\times GL(l,\cx)$.\label{Frr}\end{proposition}
\begin{proof} This is the same argument as in the proof of Proposition \ref{rat_pairs}. We can uniquely conjugate $\beta_+(0)$ to a matrix $B_+$ of the form \eqref{beta+} (resp. \eqref{beta-}) if $k<l$ (resp. $k\geq l$), with $\beta_-(0)$ being a matrix $B_-$ of the form \eqref{beta-} if $k\leq l$ and of the form  \eqref{beta+} if $k>l$, and $(f_1,\dots,f_k)=(0,\dots,0,1)$ in both cases. There is a unique pair $(g_-,g_+)$ of gauge transformations on $[-1,0]$ and $[0,1]$ with $g_\pm(0)=1$ which make $\alpha$ identically zero. Thus $g_-^{-1}(-1)B_-g_-(-1)=\beta_-(-1)$ and $g_+^{-1}(1)B_+g_+(1)=\beta_+(1)$. The desired biholomorphism is given by associating to a solution $\bigl(\alpha(t),\beta(t)\bigr)$ the invertible matrices $g_-(-1),g_+(+1)$ and the characteristic polynomials of $B_-$ and $B_+$.\end{proof}

\section{Spaces of curves and divisors\label{n+m-1}}

This section is largely technical, given to fix the notation and introduce certain notions needed later on.

\subsection{The Douady space of $\cx^2$} According to \cite{Nak} and \cite{CM},  the Douady space $\bigl(\cx^2\bigr)^{[m]}$, parameterising $0$-dimensional complex subspaces of length $m$ in $\cx^2$, can be represented by the manifold ${\sH}_{m}$ of $GL(m,\cx)$-equivalence classes of
\begin{equation}\tilde{\sH}_m  =\left\{(A,B,v)\in \gl(m,\cx)^2\times\cx^m;\enskip [A,B]=0,\enskip
 \cx^m=\text{\rm Span}\bigl\{A^iB^jv\bigr\}_{i,j\in\oN}\right\}.\label{Hilbert0}\end{equation}
The correspondence is induced by the $GL(n,\cx)$-invariant map $\tilde{\sH}_m\rightarrow\bigl(\cx^2\bigr)^{[m]}$, which assigns to
$(A,B,v)$ the complex space $Z$, the support of which are the pairs of eigenvalues of $A$ and $B$ ($A$ and $B$ commute), with $\sO_Z=\sO(U)/I$,
where $U$ is a neighbourhood of $\supp Z$ and $I$ is the kernel of the map
\begin{equation}\psi:\sO(U)\rightarrow \cx^m,\quad \psi(f)=f(A,B)v.\label{psi}\end{equation}
Let $Y\subset \bigl(\cx^2\bigr)^{[m]}\times \cx^2$ be the tautological subspace (i.e. $(Z,t)\in Y \iff t\in \supp Z$) and let $\sW_m$ be the pushdown of the
structure sheaf of $Y$ onto $\bigl(\cx^2\bigr)^{[m]}$. As a vector bundle, the fibre of $\sW_m$ at $Z\in \bigl(\cx^2\bigr)^{[m]}$ is
$H^0(Z,\sO_Z)$. Following Nakajima \cite{Nak}, we call $\sW_m$ the {\em tautological vector bundle}.
In the above matricial model, $\sW_m$ is the vector bundle associated to the principal $GL(m,\cx)$-bundle $\tilde{\sH}_m$ over $\sH_m$.
\par
The next step is to make $\sW_m$ into a Hermitian vector bundle. Given the usual correspondence between the complex quotient of the set of stable points and the K\"ahler quotient, we can identify (cf.
\cite{Nak}) $\sH_m$ with the manifold of $U(m)$-equivalence classes of
\begin{equation*}\hat{{\sH}}_m  =\left\{(A,B,v)\in \gl(m,\cx)^2\times\cx^m;\enskip [A,B]=0,\enskip
 \bigl[A,A^\ast\bigr]+\bigl[B,B^\ast]+vv^\ast=1\right\}.\label{Hilbert}\end{equation*}
The bundle $\sW_m$ is now isomorphic to $\hat{{\sH}}_m\times_{U(m)} \cx^m$ and, hence, it inherits a Hermitian metric from the standard metric on $\cx^m$. More explicitly, this metric is defined as follows.
Let $Z\in \bigl(\cx^2\bigr)^{[m]}$ be represented by $(A,B,v)$ satisfying both equations in the definition of $\hat{{\sH}}_m$, and let $\bar{f},\bar{g}\in
\sO_Z=\sO(U)/I$ be represented by $f,g\in \sO(U)$. Then:
\begin{equation} \langle \bar{f},\bar{g}\rangle= \langle f(A,B)v,g(A,B)v\rangle,\label{Hermitian}\end{equation}
where the second metric is the standard Hermitian inner product on $\cx^m$.

\subsection{The Douady space of $\TP$}
We consider now the Douady space $T^{[m]}$  of $T=\TP$, parameterising $0$-dimensional complex subspaces of length $m$ in $T$.
Recall that $T=\TP$ is obtained by glueing together two copies $U_0,U_\infty$ of $\cx^2$. According to \cite{CM}, we obtain $T^{[m]}$ by an analogous glueing of $U_0^{[m]}$ and $U_\infty^{[m]}$. 
We take two copies $\tilde{\sH}^0_m,\tilde{\sH}^\infty_m$ of \eqref{Hilbert0}, with ``coordinates" $A,B,v$ and $\tilde{A},\tilde{B},\tilde{v}$ and glue them together over the subset $\det A\neq 0\neq \det \tilde{A}$ by:
$$ \tilde{A}=A^{-1},\quad \tilde{B}=BA^{-2},\quad\tilde{v}=v.$$
Call the resulting manifold $\tilde{\sT}_m$. The glueing is $GL(m,\cx)$-equivariant and we obtain a manifold $\sT_m=\tilde{\sT}_m/GL(m,\cx)$ which represents $T^{[m]}$. The tautological bundle $\sW_m$ over $T^{[m]}$ is the vector bundle associated to the principal $GL(m,\cx)$-bundle $\tilde{\sT}_m$ over $\sT_m$.
\begin{remark} Unsurprisingly, one cannot glue together the unitary descriptions of $U_0^{[m]},U_\infty^{[m]}$. In particular, we do not have a natural Hermitian metric on $\sW_m$ over $T^{[m]}$.\end{remark}

\subsection{Curves and divisors}
Let $\sC_n$ denote the space of all curves  $S\in |\sO(2n)|$, i.e. space of polynomials of the form \eqref{S}. Thus, $\sC_n\simeq \cx^{n^2+2n}$. Let $\sY_n\subset T\times \sC_n $ be the resulting correspondence, i.e.
\begin{equation} \sY_n=\bigl\{(t,S)\in T\times \sC_n;\enskip t\in S \bigr\}.\label{sY}\end{equation}
We have the two projections: $p_1:\sY_n\rightarrow T$ and $p_2:\sY_n\rightarrow \sC_n$. We denote by $\sY_{n,m}$ the relative $m$-Douady space for $p_2:\sY_n\rightarrow \sC_n$. It is a complex space \cite{Pou} with a projection $p:\sY_{n,m}\rightarrow \sC_n$, and its points are pairs $(S,\Delta)$, where $S\in \sC_n$ and $\Delta$ is an effective Cartier divisor of degree $m$ on $S$.
There is a natural holomorphic map
\begin{equation} \phi: \sY_{n,m}\rightarrow T^{[m]},\label{phi}\end{equation}
which assigns to $(S,\Delta)$ the complex subspace $Z=(\supp\Delta,\sO_\Delta)$, where $\sO_\Delta$ is given by the ideal generated by $\Delta$ (as a Cartier divisor) and  the polynomial \eqref{S} defining $S$. 
\par
We have two canonical subsets of $\sY_{n,m}$:
\begin{equation} \sY_{n,m}^0=\bigl\{(S,\Delta);\enskip \infty\not\in\pi(\supp\Delta)\bigr\},\quad \sY_{n,m}^\infty=\bigl\{(S,\Delta);\enskip 0\not\in\pi(\supp\Delta)\bigr\}.\label{technical}\end{equation}
The map $\phi$ maps $\sY_{n,m}^0$ into $U_0^{[m]}$ and $\sY_{n,m}^\infty$ into $U_\infty^{[m]}$.

\subsection{Line bundles\label{DLB}}
Let now $E$ be a line bundle on $\TP$, the transition function of which from $U_0$ to $U_\infty$ is $\rho(\zeta,\eta)$. We fix a trivialisation of $E$ on $U_0,U_\infty$ (since $H^0(\TP,\sO)=\cx$, such a trivialisation of $E$ on $U_0,U_\infty$ is determined up to a constant factor).
\par
For any $(S,\Delta)\in
\sY_{n,m}^0$, we obtain a map 
\begin{equation} \Phi: H^0(S,E_{|S})\rightarrow  H^0(\supp\Delta,\sO_\Delta),\label{Phi_0}\end{equation}
from $H^0(S,E_{|S})$ to the fibre of $\sW_m$ over $\phi(S,\Delta)$ 
by first representing a section by a pair of holomorphic functions $s_0,s_\infty$ on $U_0\cap
S$, $U_\infty\cap S$, satisfying
 $s_\infty=\rho s_0$ on $U_0\cap U_\infty\cap S$, and taking an extension of $s_0$ to some neighbourhood $U$ of $U_0\cap S$ in $U_0$.
 \par
If we denote by $\sE$ the linear space over $\sY_{n,m}$, the fibre of which over $(S,\Delta)$ is $H^0(S,E_{|S})$ (i.e. $\sE$ is the pullback of the analogously defined linear space over $\sC_n$), then $\Phi$ makes the following
diagram commute:
\begin{equation}\begin{CD} \sE @>\Phi >> \sW_m\\ @VVV  @VVV \\ \sY_{n,m}^0 @>\phi >> U_0^{[m]} . \end{CD}
\label{Phi}\end{equation}
Obviously the above discussion holds for $\sY_{n,m}^\infty$ as well.

 We now specialise to the case $E=F(n+p-1)$, where $F$ is a line bundle  on $\TP$ with $c_1(F)=0$. Let $S\in |\sO(2n)|$ be of the form \eqref{S}, and let
$\Delta$ be an effective divisor on $S$ of degree $pn$ such that
\begin{equation} H^0\bigl(S,F(n+p-2)[-\Delta]\bigr)=0.\label{nosecs}\end{equation}
 Let $\zeta_0\in \oP^1-\pi(\supp\Delta)$ and $D_{\zeta_0}=S\cap (\zeta-\zeta_0)$ be the divisor of points lying over $\zeta_0$. We write
\begin{equation}\begin{array}{l} V=H^0\bigl(S,F(n+p-1)\bigr)\\ V_\Delta=H^0\bigl(S,F(n+p-1)[-\Delta]\bigr)\\
V_{\zeta_0}=H^0\bigl(S,F(n+p-1)[-D_{\zeta_0}]\bigr).\end{array}\label{Vs} \end{equation} The condition \eqref{nosecs} and the fact that
$F(n+p-2)[-\Delta]$ has degree equal to $\text{genus}(S)-1$ imply that the first cohomology of $F(n+p-2)[-\Delta]$ vanishes. Therefore, the
first cohomology of $F(n+p-2)$ and of $F(n+p-1)$ vanish as well. Consequently $\dim V=np+n$ and $\dim V_\Delta=n$. Since
$[D_{\zeta_0}]=\sO_S(1)$, $\dim V_{\zeta_0}=np$, and
$$H^0\bigl(S,F(n+p-1)[-\Delta-D_{\zeta_0}]\bigr)=H^0\bigl(S,F(n+p-2)[-\Delta]\bigr)=0,$$
we have that
\begin{equation} V=V_\Delta\oplus V_{\zeta_0}.\label{decompose}\end{equation}
Moreover, we have an isomorphism:
\begin{equation} V_{\zeta_0}\longrightarrow H^0\bigl(\supp\Delta,F(n+p-1)[-D_{\zeta_0}]\bigr).\label{iso}\end{equation}
\begin{definition} We write $\sY_{n,pn}(F)$ for the subset  of $\sY_{n,pn}$ on which \eqref{nosecs} is satisfied.
\par
If $\zeta_0\in \oP^1$, then we write $\sY_{n,pn}(\zeta_0)$ for the subset  of $\sY_{n,pn}$ on which $\zeta_0\not\in \pi(\supp\Delta)$. We also write $\sY_{n,pn}(F,\zeta_0)=\sY_{n,pn}(\zeta_0)\cap \sY_{n,pn}(F)$ and we use the superscripts $0,\infty$ to denote the intersections of any of these sets with 
$\sY_{n,pn}^0$ or $\sY_{n,pn}^\infty$.
\par
We write $\sV,\sV_\Delta,\sV_{\zeta_0}$ for the vector bundles over $\sY_{n,pn}(\zeta_0)$, the fibres of which over $(S,\Delta)$ are, respectively, the vector spaces $V,V_\Delta,V_{\zeta_0}$, given by \eqref{Vs}. 
\end{definition}

If $\zeta_0\neq \infty$, then the isomorphism \eqref{iso} can be interpreted as the top map in \eqref{Phi} for $E=F(n+p-1)[-D_{\zeta_0}]$. In particular, we obtain a Hermitian metric on $\sV_{\zeta_0}$ over $\sY_{n,pn}^0(F,\zeta_0)$. 
Similarly, if  $\zeta_0\neq 0$, then we obtain a Hermitian metric on $\sV_{\zeta_0}$ over $\sY_{n,pn}^\infty(F,\zeta_0)$.
\par
We finally specialise to the case $F=L^t$ and we write, for any interval $I$:
\begin{equation} \sY_{n,pn}(I)=\bigcap_{t\in I} \sY_{n,pn}(L^t).\label{Ii}\end{equation}
The notation $\sY_{n,pn}(I,\zeta_0)$, $\sY_{n,pn}^0(I,\zeta_0)$ and $\sY_{n,pn}^\infty(F,\zeta_0)$ is then self-explanatory.

\subsection{Translations\label{trans}} Let $c(\zeta)$ be a quadratic polynomial, viewed as a section of $\pi^\ast\sO(2)$ on $T$. It induces a fibrewise translation on $T$:
$$ (\zeta,\eta)\mapsto \bigl(\zeta,\eta+c(\zeta)\bigr),$$
which in turn induces a translation $t_{c(\zeta)}:\sY_{n,m}\rightarrow \sY_{n,m}$. We have a similar map on $T^{[m]}$, given by
\begin{equation}\tilde{\sH}_m\ni (A,B)\mapsto \bigl(A,B+c(A)\bigr)\in \tilde{\sH}_m. \label{tc1}\end{equation}
We denote this map also by $t_{c(\zeta)}$. The following diagram commutes
\begin{equation}\begin{CD}  \sY_{n,m} @>\phi >> T^{[m]}\\ @V t_{c(\zeta)}VV  @VVt_{c(\zeta)}V \\ \sY_{n,m} @>\phi >> T^{[m]} . \end{CD}
\label{tc2}\end{equation}
The formula \eqref{tc1} defines a map on the tautological bundle $\sW_m$ over $T^{[m]}$. In terms of $\sO_Z$, $Z$ being a $0$-dimensional subspace of length $m$, this map is given by
\begin{equation} f(\zeta,\eta)\mapsto f\bigl(\zeta,\eta+c(\zeta)\bigr).\label{tc3}\end{equation}
We remark that this last map is not an isometry over $U_0^{[m]}$ or over $U_\infty^{[m]}$.

\section{Asymptotics of curves}

In this section, we consider the asymptotic behaviour of two spectral curves, the centres of which move away from each other. We define first an
$SO(3)$-invariant distance function between curves in $\sC_n$.  On $\oP^1$ distance is measured in the standard round Riemannian metric of
diameter $\pi$ on $S^2$. This induces a fibrewise inner product on $\TP$. Let $d_H$ be the induced fibrewise Hausdorff distance between
sets and $\pi:\TP\rightarrow \oP^1$ be the projection. For two curves $S,S^\prime$ in $|\sO(2n)|$ we define their distance $d(S,S^\prime)$
by
\begin{equation} d(S,S^\prime)=\max\left\{ d_H\bigl(S\cap \pi^{-1}(w), S^\prime\cap \pi^{-1}(w)\bigr); w\in S^2\right\}. \label{distance}\end{equation}
 The distance $d$ is equivalent to the supremum of the Euclidean distance between roots of the polynomials \eqref{S} defining $S,S^\prime$ as we vary
 $\zeta$ over a relatively compact open set.
 \par
For a curve $S\in \sC_n$, given in $U_0$ by the equation
$$\eta^n+a_1(\zeta)\eta^{n-1}+\cdots +a_{n-1}(\zeta)\eta+ a_n(\zeta)=0,$$
we define its {\em centre} as
\begin{equation}c(\zeta)=a_1(\zeta)/n.\label{centers}\end{equation}
In addition, we set
\begin{equation}C(S)=\{(\zeta,\eta); (\eta+ c(\zeta))^{n}=0\}.\label{center}\end{equation}
\par
We shall consider next a pair of real curves $S_1\in |\sO(2k)|$ and $S_2\in |\sO(2l)|$. Let $c_1(\zeta),c_2(\zeta)$ be their  centres.
These are quadratic polynomials invariant under the antipodal map, and we write
$$ c_1(\zeta)=z_1+2x_1\zeta-\bar{z}_1\zeta^2,\quad c_2(\zeta)=z_2+2x_2\zeta-\bar{z}_2\zeta^2.$$
 Let
\begin{equation} R=R(S_1,S_2)=\sqrt{(x_1-x_2)^2+|z_1-z_2|^2}\label{r}\end{equation} be the distance
between the centres and let
\begin{equation} \zeta_{12}=\frac{x_1-x_2+R}{\bar{z}_1-\bar{z}_2} \enskip\text{and}\enskip \zeta_{21}=\frac{x_1-x_2-R}{\bar{z}_1-\bar{z}_2}
\label{zeta12}\end{equation} be the two intersection points of the polynomials $c_1(\zeta)$ and $c_2(\zeta)$, i.e. the two opposite
directions between the centres. Recall that $S_1\cap S_2$ denotes a complex subspace of $T$, and, in an appropriate context, a Cartier divisor on $S_1$ or $S_2$.
\par
Recall the set $\sS_{k,l}$ of pairs of curves (plus a choice of a divisor $D$) defined in \ref{Skl}.
For every $K>0$ we define the following region of $\sS_{k,l}$:
\begin{equation}\sS_{k,l}(K)=\left\{ (S_1,S_2)\in \sS_{k,l} ;\enskip d\bigl(S_i,C(S_i)\bigr)\leq K, \enskip i=1,2,\right\}.\label{K}\end{equation}
\par

A priori, we do not know that $\sS_{k,l}(K)$ has nonempty interior (it could happen that, when $R\rightarrow \infty$, then $d\bigl(S_i,C(S_i)\bigr)\rightarrow 0$). We shall prove that it is so. First of all, we have
\begin{lemma} Let $c_1(\zeta)$ and $c_2(\zeta)$ be two quadratic polynomials, invariant under the antipodal map. Then the pair of curves defined by $(\eta+ c_1(\zeta))^{k}=0$ and $(\eta+ c_2(\zeta))^{l}=0$ belongs to $\sS_{k,l}$.\label{points}\end{lemma}
\begin{proof} One needs to show that there exists a solution to Nahm's equations on $[-1,0)\cup(0,1]$ with the correct matching conditions (those of $N_{k,l}$) at $t=0$, and such that the corresponding spectral curves are the given ones. We can, in fact, find it on $(-\infty,0)\cup (0,+\infty)$. We observe that such a solution is a point in the hyperk\"ahler quotient of $F_{k,l}(-1,1)\times O_k\times O(l)$ by $U(k)\times U(l)$, where $O_k$ and $O_l$ are regular nilpotent adjoint orbits in $\gl(k,\cx)$ and $\gl(l,\cx)$ with Kronheimer's metric \cite{Kron} and $F_{k,l}(-1,1)$ was defined in Remark \ref{Fkl}. One shows, as in \cite{CMP2} (using nilpotent orbits, rather than the semi-simple ones) that this hyperk\"ahler quotient is a one-point set.  \end{proof}

The proof shows that a solution to Nahm's equations, corresponding to this pair of curves, exists on $(-\infty,0)\cup (0,+\infty)$. Its restriction to $[-1,0)\cup(0,1]$ defines an element of $F_{k,l}^{\rm rr}$, as long as $c_1(\zeta)\neq c_2(\zeta)$. Let $(v_-^0,g_-^0,v_+^0,g_+^0)$  be the corresponding element of $\cx^k\times GL(k,\cx)\times \cx^l\times GL(l,\cx)$, given by Proposition \ref{Frr}. Observe that $v_-$ and $v_+$ are the coefficients of polynomials $(\eta+ c_1(0))^{k}$ and $(\eta+ c_2(0))^{l}$.

\begin{proposition}  For any $L>0$, there exists a $K=K(L,k,l)>0$ with the following property. Let $c_i(\zeta)=z_i+2x_i\zeta-\bar{z}_i\zeta^2$, $i=1,2$, and suppose that $|z_1-z_2|\geq 1$. Let
$(v_-,g_-,v_+,g_+)\in \cx^k\times GL(k,\cx)\times \cx^l\times GL(l,\cx)$ and let $q_-(z)$ and $q_+(z)$ be polynomials, the coefficients of which are given by the entries of $v_-$ and $v_+$, so that $q_-(z),q_+(z)$ are the characteristic polynomials of $B_-,B_+$, defined in the proof of Proposition \ref{Frr}. Suppose that all  roots of $q_-(z)$ (resp. roots of $q_+(z)$) satisfy $|r-c_1(0)|\leq L$ (resp. $|r-c_2(0)|\leq L$) and that 
\begin{equation} \|\ln g_-^\ast g_- -\ln (g_-^0)^\ast g_-^0\|\leq 2L,\quad \|\ln g_+^\ast g_+ - \ln (g_+^0)^\ast g_+^0\|\leq 2L\label{LL}\end{equation} 
(here $\ln$ denotes the inverse to the exponential mapping restricted to hermitian matrices).   Then the pair of spectral curves corresponding, via Proposition \ref{Frr}, to  $(v_-,g_-,v_+,g_+)$  lies in $\sS_{k,l}(K)$.\label{L-K}\end{proposition}
\begin{proof}  Let $r^1,\dots,r^k$ (resp. $s^1,\dots,s^l$ be the roots of $q_-(z)$ (resp. $q_+(z)$). Consider a solution to Nahm's equations on $(-\infty,0)\cup (0,+\infty)$, with the correct matching conditions at $t=0$, and such that the corresponding pair of spectral curves is $\prod_i \bigl(\eta+r^i+2x_1\zeta-\overline{r^i}\zeta^2\bigr)=0$ and $\prod_i \bigl(\eta+s^i+2x_2\zeta-\overline{s^i}\zeta^2\bigr)=0$. Such a solution exists just as the one in Lemma \ref{points} (this follows directly from \cite{CMP2}). Its restriction to  $[-1,0)\cup(0,1]$ defines an element of $F_{k,l}^{\rm rr}$ and the proofs in \cite{CMP2} show that the corresponding $g_-^1,g_+^1$ satisfy the estimate \eqref{LL}. Let $\bigl((\alpha_-,\alpha_+),(\beta_-,\beta_+)\bigr)$ be this solution to Nahm's equations. Moreover, the estimates of Kronheimer \cite{Kr} and Biquard \cite{Biq} show that for $t\leq -1/2$ and $t\geq 1/2$ the solution to Nahm's equations is within some $C$ from its centre (i.e. $T_i(t)$ are within distance $C$ from theirs centres for $i=1,2,3$). Let $h_-=g_-\bigl(g_-^1\bigr)^{-1}$ and $h_+=g_+\bigl(g_+^1\bigr)^{-1}$ and let $h_-(t)$ (resp. $h_+(t)$) be a path in $GL(k,\cx)$ (resp. $GL(l,\cx)$) with $h_-(-1)=h_-$ and $h_-(t)=1$ for $t\in [-1/2,0]$ (resp.  $h_+(1)=h_+$ and $h_+(t)=1$ for $t\in [0,1/2]$). Define a solution to the {\em complex} Nahm equation by acting on $\bigl((\alpha_-,\alpha_+),(\beta_-,\beta_+)\bigr)$ with the complex gauge transformations $h_\pm(t)$. If we now solve the real Nahm equation via a complex gauge transformation $G(t)$, which is identity at $\pm 1$, then the corresponding element of $\cx^k\times GL(k,\cx)\times \cx^l\times GL(l,\cx)$ is the given one. On the other hand, the left-hand side of the real Nahm equation is bounded, because $\beta_\pm (t)$ and $\bigl(\alpha_\pm(t)+\alpha_\pm(t)^\ast\bigr)/2$ are within $C$ from their centres for $t\in [-1,-1/2]\cup [1/2,1]$. Then it follows from estimates of Donaldson and Hurtubise (see section 2 in \cite{Hu}) that the hermitian part of $\dot{G}G^{-1}$ is uniformly bounded at $t=\pm 1$, which proves the estimate ($K$ is determined by $C$ and the bound on $\dot{G}G^{-1}(\pm 1)$.
\end{proof}

As a corollary (of the proof) we can give an estimate on spectral curves of clusters in terms of the corresponding rational map:

\begin{corollary} For any $L>0$, there exists a $K=K(L,k,l)>0$ with the following property. Let $\left(\frac{p_1(z)}{q_1(z)},\frac{p_2(z)}{q_2(z)}\right)\in {\rm Rat}_k\bigl(\oP^1\bigr)\times {\rm Rat}_l\bigl(\oP^1\bigr)$ be a pair of rational maps and let $\beta_1^1,\dots,\beta_k^1$ (resp.  $\beta_1^2,\dots,\beta_l^2$) be the roots of $q_1(z)$ (resp. $q_2(z)$). Suppose that the functions satisfy:
\begin{itemize}
\item[(i)] $|\beta_i^1-\beta_j^2|\geq 1$ for any $i,j$.
\item[(ii)] $|\beta_i^s-\beta_j^s|\leq 2L$ for any $i,j$ and $s=1,2$.
\item[(iii)] $\bigl|\ln |p_s(\beta_i^s)|-\ln |p_s(\beta_j^s)|\bigr|\leq 2L$ for any $i,j$ and $s=1,2$.
\end{itemize}
Let $(S_1,S_2)\in \Sigma_{k,l}$ correspond to the above pair of rational functions via Proposition \ref{rat_pairs}. Then $(S_1,S_2)\in \sS_{k,l}(K)$. Moreover, if $b_1=\sum \beta_i^1/k$, $b_2=\sum \beta_i^2/l$, $a_1=\sum \ln|p_1(\beta_i^1)| /2k$, $a_2=\sum \ln|p_2(\beta_i^2)| /2l$, then $|b_s-z_s|\leq K$, $|a_s-y_s|\leq K$, $s=1,2$, where $z_s+2y_s\zeta-\bar{z}_s\zeta^2$ is the centre of $S_s$.
\hfill $\Box$\label{eest}\end{corollary}
\begin{proof} Once again consider the solution $\bigl((\alpha_-,\alpha_+),(\beta_-,\beta_+)\bigr)$ to Nahm's equations on $[-1,0)\cup (0,1]$ with $r^i=\beta_i^1$, $i=1,\dots,k$, $s^j=\beta^2_j$, $j=1,\dots,l$, $x_s=a_s$, $s=1,2$. The assumption (i) and Kronheimer's estimates \cite{Kr} imply that, near $t=\pm 1$, the solution is within some constant $C$ from the diagonal one (after acting by $U(k)$ and $U(l)$), and that the derivatives of the solution are bounded by $C$. Let us act by a complex gauge transformation, which differs from the identity only near $\pm 1$ and which diagonalises there $\beta_\pm$. We also require that $\alpha_\pm$ becomes diagonal near $\pm 1$ and that after extending this solution to the complex Nahm equation to  $[-2,0)\cup (0,2]$ by symmetry, it corresponds, via Proposition \ref{rat_pairs} to the given pair of rational maps. The remainder of the proof follows that of the previous proposition word by word.
\end{proof}

We observe that if $(S_1,S_2)\in \sS_{k,l}(K)$ and $p\in \supp S_1\cap S_2$, then $\pi(p)$ is within $b(K)/R$ from either $\zeta_{12}$ or from
$\zeta_{21}$ for some function $b(K)$.  We would like to argue that $\pi(p)$ must lie within $b(K)/R$ from $\zeta_{21}$, but we can only prove a somewhat weaker result:
\begin{proposition} For every $L>0$ and $\delta>0$, there is an $R_0$ with the following property. Let  $(S_1,S_2)\in \sS_{k,l} $ be obtained from a $(v_-,g_-,v_+,g_+)\in \cx^k\times GL(k,\cx)\times \cx^l\times GL(l,\cx)$, which satisfies the assumptions of Proposition \ref{L-K} and suppose, in addition, that $R(S_1,S_2)\geq R_0$. Then 
 the divisor $D\subset S_1\cap S_2$ may
be chosen so that $\pi(\supp D)$ is within distance $\delta $ from the point $\zeta_{21}$. \label{good_D}
\end{proposition}
\begin{proof} First of all, observe that the subset of $\sS_{k,l} $ described in the statement is connected, since the corresponding subset of $\cx^k\times GL(k,\cx)\times \cx^l\times GL(l,\cx)$ is. Therefore, it is enough to show that there is $(S_1,S_2) $ in this subset such that $\pi(\supp D)$ is within some small distance, say $1$, from $\zeta_{21}$. For this we take again a pair of completely reducible curves and consider the corresponding Nahm flow as in \cite{CMP2}. The divisor $D$ can be read off a solution to Nahm's equations as in \cite{HuMu}, pp. 73--76. This,  together with a more explicit description of solutions for reducible curves, given in Sections 5, 6 and 7 of \cite{CMP2} (in particular, the formula 6.10 together with Lemma 9.6 of that paper) shows that $D$ (which is now a Weil divisor) can be chosen as those points of $S_1\cap S_2$ which are
closer to $\zeta_{21}$ than to $\zeta_{12}$ (a word of warning: the Nahm equations in \cite{HuMu} have a different sign, corresponding to the change $t\rightarrow -t$).\end{proof}

\par
We now give a picture of the asymptotic behaviour of curves in $\Sigma_{k,l}$, analogous to that of monopole spectral curves given in
\cite{AtHi}, Propositions 3.8 and 3.10. Before stating the result, we need to define an appropriate subset of $\Sigma_{k,l}$. 
\begin{definition} We denote by $\Sigma_{k,l}(K)$ the subset of $\Sigma_{k,l}\cap \sS_{k,l}(K)$ defined as follows. If $\pi(S_1\cap S_2)$ is within distance $1$ from $\{\zeta_{12},\zeta_{21}\}$, then $(S_1,S_2)\in \Sigma_{k,l}(K)$ if and only if $D$ can be chosen so that $\pi(\supp D)$ is within distance $1 $ from the point $\zeta_{21}$.\label{Sigma(K)}\end{definition}

\begin{remark} Proposition \ref{good_D}  imply that curves corresponding to rational maps satisfying the assumption of Corollary \ref{eest} belong to $\Sigma_{k,l}(K)$.\label{good_rat}\end{remark}

\begin{proposition} Let $(S_1^n,S_2^n)$ be a sequence of points in $\Sigma_{k,l}(K)$ such that the distances $R_n$ between the centres of $S_1^n$ and $S^n_2$
tend to infinity. Let $P_1^n(\zeta,\eta)=0$ and $P_2^n(\zeta,\eta)=0$ be the equations defining $S_1^n$ and $S^n_2$ and $c_1^n(\zeta)$,
$c_2^n(\zeta)$ the centres of $S_1^n$ and $S_2^n$. Then the centred curves $P_1^n\bigl(\zeta,\eta-c_1^n(\zeta)\bigr)=0$,
$P_2^n\bigl(\zeta,\eta-c_2^n(\zeta)\bigr)=0$ have a subsequence converging to spectral curves 
 of monopoles of charge
$k$ and $l$, respectively.\label{convergence}\end{proposition}
\begin{proof}
 We prove this for $S_2^n$. The centred curves, given by the polynomials $P_2^n\bigl(\zeta,\eta-c_2^n(\zeta)\bigr)=0$, lie in a compact
 subset, and so we can find a subsequence converging to some $S_2^\infty$. Let $R_n=R(S_1^n,S_2^n)$. The divisor of $P_1^n$ on $S_2^n$ is the same as that of $P^n_1/(R_n)^k$. The latter has a subsequence convergent to $c(\zeta)^k$, where $c(\zeta)$ is a quadratic polynomial. Write $\zeta_{12}$ and $\zeta_{21}$ for its roots, as in \eqref{zeta12}. Proposition \ref{good_D} implies that  the translated divisors
 $\Delta_n=\{(\zeta,\eta); (\zeta,\eta-c_2^n(\zeta))\in D_n\}$ converge to  $kD_{\zeta_{21}}$ on $S_2^\infty$ (recall that $D_{\zeta_0}$ denotes the divisor of $(\zeta-\zeta_0)$). Consider now the corresponding solutions to Nahm's equations, given by Proposition \ref{symmNahm}. The solutions
 shifted by the centres will have a convergent subsequence on $(0,2)$, thanks to Proposition 1.3 in \cite{BielAGAG}. Therefore, the sections of $L^t(k+l-1)[-\Delta_n]$ converge to sections of a line bundle over $S_2^\infty$. This line bundle must be $L^t(k+l-1)[-kD_{\zeta_{21}}]\simeq L^t(k-1)$, and, hence, the limit Nahm flow corresponds to $L^t(k-1)$. Since the limit flow is nonsingular, 
$H^0\bigl(S_2^\infty,L^t(l-2)\bigr)=0$ for $t\in(0,2)$. 
 In addition, if the Nahm matrices were symmetric at $t=1$ for $S_2^n$, then they are symmetric for $S_2^\infty$, and, hence, $L^2$ is trivial on $S_2^\infty$. Finally, $S_2^\infty$ does not have a multiple component, thanks to Remark \ref{multiple}.  \end{proof}

The proof shows that the divisors $D_n$ and $\tau(D_n)$, translated by the centres, converge as well. 
 Observe, that we have embeddings $\sS_{k,l}\hookrightarrow \sY_{k,kl}\bigl((0,2)\bigr)$ and
$\sS_{k,l}\hookrightarrow \sY_{l,kl}\bigl((0,2)\bigr)$ (recall \eqref{Ii}), given by
\begin{equation} (S_1,S_2)\mapsto (S_1,\tau(D))\in \sY_{k,kl},\quad (S_1,S_2)\mapsto (S_2,D)\in \sY_{l,kl}.\label{S->A}\end{equation}
From the proof of the above proposition, we have:
\begin{corollary} Let $\Xi_1(K)$ (resp. $\Xi_2(K)$) be the subset of $\Sigma_{k,l}(K)$ defined by $c_1(\zeta)=0$ (resp. $c_2(\zeta)=0$) and $R\geq 1$. Then
$\Xi_1(K)$ is a relatively compact subset of $\sY_{k,kl}\bigl((0,2)\bigr)$ and $\Xi_2(K)$ is a relatively compact subset of $\sY_{l,lk}\bigl((0,2)\bigr)$.
\label{compact}\end{corollary}

We also have:
\begin{corollary} There exists an $R_0$, such that, for all $(S_1,S_2)\in\Sigma_{k,l}(K)$ with $R(S_1,S_2)\geq R_0$, neither $S_1$ nor $S_2$ has multiple components. \label{nomult}\end{corollary}
\begin{proof} If this were not the case, then the limit curves obtained in Proposition \eqref{convergence} would also have a multiple component, and could not be spectral curves of monopoles.
\end{proof}

\section{Asymptotics of matricial polynomials\label{asy-pol}}

We shall now consider the flow $L^t(k+l-1)$ on $S_1\cup S_2$ for $(S_1,S_2)\in \Sigma_{k,l}$ (defined in \ref{Sigma(K)}). Observe that the corresponding matricial flow $A(t,\zeta)$ has poles at $t=0$ corresponding to the irreducible representation of dimension $k+l$, and so the boundary behaviour of $SU(2)$-monopoles. Of course, it does not have the correct boundary behaviour at $t=2$, but we are going to show that, in the
asymptotic region of $\Sigma_{k,l}(K)\subset \Sigma_{k,l}\cap \sS_{k,l}(K)$, the corresponding matricial flow is exponentially close to the block-diagonal matricial flow
corresponding to $L^t(k+l-1)[-\tau(D)]$ on $S_1$ and $L^t(k+l-1)[-D]$ on $S_2$. In particular, it is exponentially close to being symmetric at $t=1$, and so we can construct an exponentially approximate solution to Nahm's equations with the correct (monopole-like) boundary behaviour by taking $A(2-t,\zeta)^T$ on $[1,2)$.
\par
We are going to prove
\begin{theorem} For every $K>0,\delta>0$, there exist an $R_0, \alpha>0, C>0$ such that for any $(S_1,S_2)\in \Sigma_{k,l}(K)$ with $R(S_1,S_2)\geq R_0$ the
following assertions hold
\begin{itemize}
\item[1.] The line bundle $L^t(k+l-2)$ on $S_1\cup S_2$ does not lie in the theta divisor for any $t\in (0,2)$.
\item[2.] For any $t\in[\delta,2-\delta]$, the line bundle $L^t(k+l-1)$ can be represented by a
matricial polynomial $A(t,\zeta)=\bigl(T_2(t)+iT_3(t)\bigr)+2T_1(t)\zeta +\bigl(T_2(t)-iT_3(t)\bigr)\zeta^2$ such that the matrices are skew-hermitian and the
$T_i(t)$, $i=1,2,3$, are $Ce^{- \alpha R}$-close to block-diagonal skew-hermitian matrices $\hat{T}_i(t)$ with blocks defining a given
matrix-polynomial representation of $L^t(k+l-1)[-\tau(D)]$
 on $S_1$ and $L^t(k+l-1)[-D]$ on $S_2$.
\end{itemize}
\label{hardest}
\end{theorem}
The second part of the theorem can be strengthened. Let us write 
$$\hat{A}(t,\zeta)=\bigl(\hat{T}_2(t)+i\hat{T}_3(t)\bigr)+2\hat{T}_1(t)\zeta +\bigl(\hat{T}_2(t)-i\hat{T}_3(t)\bigr)\zeta^2.$$
\begin{theorem} With the notation and assumptions of the previous theorem, there exists a map $g:[\delta,2-\delta]\times \oP^1\rightarrow SL(k+l,\cx)$, analytic in the first variable and meromorphic in the second variable, such that $g(t,\zeta)A(t,\zeta)g(t,\zeta)^{-1}=\hat{A}(t,\zeta)$, for any $(t,\zeta)\in [\delta,2-\delta]\times \oP^1$.
Moreover:
\begin{itemize}
\item[(i)] There are constants $C,\alpha>0$ such that, for any $t\in[\delta,2-\delta]$ and any $\zeta_1,\zeta_2\in \oP^1$ with $|\zeta_i- \zeta_{12}|\geq 1/2$, $|\zeta_i- \zeta_{21}|\geq 1/2$, $i=1,2$, $\|g(t,\zeta_1)g(t,\zeta_2)^{-1}-1\|\leq Ce^{- \alpha R}$ (as matrices).
\item[(ii)] If we write $g$ in the block form as $\left(\begin{array}{cc} g_{11} & g_{12}\\ g_{21} & g_{22}\end{array}\right)$, with $g_{11}$ being $k\times k$ and $g_{22}$ $l\times l$, then the only poles of $g_{11}(t,\zeta)$ and $g_{12}(t,\zeta)$ may occur at $\zeta\in \pi \bigl(\supp \tau(D)\bigr)$ and the only poles of $g_{21}(t,\zeta)$ and $g_{22}(t,\zeta)$ may occur at $\zeta\in \pi(\supp D)$.
\end{itemize}\label{hardest2}\end{theorem}

The remainder of the section is taken by a proof of these theorems.
\par
{\bf Step 1.} 
 Let $P_1(\zeta,\eta)=0$, $P_2(\zeta,\eta)=0$  be the equations of
$S_1$ and $S_2$. Let $c_1(\zeta),c_2(\zeta)$ be the centres of $S_1,S_2$ (defined by \eqref{centers}). Consider the effect of shifting the
curves by the ``total centre" $c_{12}=\frac{k}{k+l}c_1+ \frac{l}{k+l}c_2$, i.e. curves defined by $P_1(\zeta,\eta-c_{12}(\zeta))=0$,
$P_2(\zeta,\eta-c_{12}(\zeta))=0$. The effect is the same on matrices $\hat{T}_i$ and $T_i$: adding a matrix in the centre of $U(k+l)$.
Thus, we can assume, without loss of generality, that $c_{12}(\zeta)=0$, i.e. that the centres of curves $S_1,S_2$ satisfy
\begin{equation} kc_1(\zeta)+lc_2(\zeta)=0.\label{central}\end{equation}
We can also assume, using the $SO(3)$-action, that $\zeta_{21}=0$ (recall \eqref{zeta12}). This means that the centre of $S_1$ is
$\frac{lR\zeta}{k+l}$  ($R=R(S_1,S_2)$), and the centre of $S_2$ is $-\frac{kR\zeta}{k+l}$. Finally, thanks
to Proposition \ref{good_D}, we can take $R_0$ large enough, so that $\pi(\supp D)\subset \overline{B(0,1/2)}$. Choose now a $\zeta_0\in \oP^1$ with $d(\zeta_0,0)>1/2$ and $d(\zeta_0,\infty)>1/2$.  Following \eqref{Vs}, write
$$V^i(t)=H^0\bigl(S_i,L^t(k+l-1)\bigr),\enskip V^i_{\zeta_0}(t)=H^0\bigl(S_i,L^t(k+l-1)[-D_{\zeta_0}]\bigr),\enskip i=1,2,$$
$$V^1_\Delta(t)=H^0\bigl(S_1,L^t(k+l-1)[-\tau(D)]\bigr), \enskip V^2_\Delta(t)=H^0\bigl(S_2,L^t(k+l-1)[-D]\bigr).$$
For $t\in (0,2)$, we have the decompositions \eqref{decompose}:
$$ V^i(t)=V^i_\Delta(t)\oplus V^i_{\zeta_0}(t),\enskip i=1,2.$$
The idea of the proof is that sections of $V^1(t)$ and $V^2(t)$, which are, in this decomposition, of the form $s+0$ ($s\in V^i_\Delta(t)$),
are exponentially close (in a sense to be defined) to sections of $L^t(k+l-1)$ on $S_1\cup S_2$.
\par
{\bf Step 2.}
We now consider arbitrary curves and divisors, as in Section \ref{n+m-1}.
Recall, from Section \ref{DLB}, the vector bundles $\sV^i(t), \sV^i_\Delta(t), \sV^i_{\zeta_0}(t)$ over $\sY_{k,kl}(\zeta_0)$ and $\sY_{l,kl}(\zeta_0)$, the fibre of which at $(S_1,S_2)$ are
$V^i(t), V^i_\Delta(t), V^i_{\zeta_0}(t)$. We denote by the same letters the corresponding vector bundles over $\sS_{k,l}$ or, rather, over the subset $\sS_{k,l}(\zeta_0)$, on which $\zeta_0\not\in\pi(\supp S_1\cap S_2)$. We shall usually not write this $\zeta_0$, keeping in mind, that it should be inserted wherever  $V^i_{\zeta_0}(t)$ is discussed.

There are embeddings $\lambda_{11},\lambda_{12}:\sS_{k,l}\rightarrow \sY_{k,kl}$ and $\lambda_{21},\lambda_{22}:\sS_{k,l}\rightarrow \sY_{l,kl}$ (cf. \eqref{S->A}):
\begin{equation}\lambda_{ij}(S_1,S_2)=\bigl(S_j,\tau^i(D)\bigr),\enskip i,j=1,2\label{embeds}\end{equation}
(recall that $\tau^2=\text{Id}$). Observe that $\lambda_{11}$ maps into $\sY_{k,kl}^\infty$, $\lambda_{12}$ into $\sY_{l,kl}^\infty$, $\lambda_{21}$ into $\sY_{l,kl}^0$ and $\lambda_{22}$ into $\sY_{l,kl}^0$. We have the maps $\Phi_{ij}$, $i,j=1,2$, defined as follow:  $\Phi_{11}$ is the top map in \eqref{Phi} over $\sY^\infty_{k,kl}$ for $\sE= \sV^1_{\zeta_0}(t)$, $\Phi_{21}$ is the top map in \eqref{Phi} over $\sY^0_{k,kl}$ for $\sE= \sV^1_{\zeta_0}(t)$, $\Phi_{12}$ is the top map in \eqref{Phi} over $\sY^\infty_{l,kl}$ for $\sE= \sV^2_{\zeta_0}(t)$, and, finally, $\Phi_{22}$ is the top map in \eqref{Phi} over $\sY^0_{l,kl}$ for $\sE= \sV^2_{\zeta_0}(t)$. We have the corresponding maps $\Lambda_{ij}$ for the bundles $\sV^j_{\zeta_0}(t)$ over $\sS_{k,l}$.
\par
A section of 
$L^t(k+l-1)$ on $S_1\cup S_2$ corresponds to a pair of sections $s_1\in H^0\bigl(S_1,L^t(k+l-1)\bigr)$, $s_2\in
H^0\bigl(S_2,L^t(k+l-1)\bigr)$ such that
\begin{equation} \Lambda_{11}(s_1)=\Lambda_{12}(s_2),\quad
\Lambda_{21}(s_1)=\Lambda_{22}(s_2).\label{Lambda}\end{equation}
We shall want to write these equations in terms of bases. 
Recall, from Corollary \ref{compact}, the subsets $\Xi_1(K)$  and $\Xi_2(K)$ of $\Sigma_{k,l}(K)$. The argument in the proof of Proposition \ref{convergence} shows that $\lambda_{ij}(\Xi_j(K))$ are relatively compact sets for $i,j=1,2$. We write $\Xi_{ij}(K)$ for the compact sets $\overline{\lambda_{ij}(\Xi_j(K))}$.
Corollary \ref{compact} says that $\Xi_{11}(K)$ (resp. $\Xi_{22}(K)$) is actually a subset of $\sY_{k,kl}^\infty\bigl((0,2)\bigr)$ (resp. a subset of $\sY_{l,kl}^0\bigl((0,2)\bigr)$). Recall, from the end of Section \ref{DLB}, that the bundles $\sV^1_{\zeta_0}(t)$ over $\sY_{k,kl}^\infty\bigl((0,2)\bigr)$ and $\sV^2_{\zeta_0}(t)$ over $\sY_{l,kl}^0\bigl((0,2)\bigr)$ have Hermitian metrics induced by maps $\Phi_{11}$ and $\Phi_{22}$. These gives us Hermitian metrics on $\sV^j_{\zeta_0}(t)$, $j=1,2$, over $\Sigma_{k,l}$. In other words, we choose Hermitian metrics on these bundles which make $\Lambda_{11}$ and $\Lambda_{22}$ isometric.
Since $\Xi_{11}(K)$ and $\Xi_{22}(K)$ are compact, there exists a constant $M=M(K,t)$, such that any vector $s_1$ of length one in the restriction of $\sV^1_{\zeta_0}(t)$ to $\Xi_{11}(K)$ and any vector $s_2$ of length one in the restriction of $\sV^2_{\zeta_0}(t)$ to $\Xi_{22}(K)$ satisfies:
\begin{equation} \left|\Phi_{21}(s_1)\right|\leq M,\quad \left|\Phi_{12}(s_2)\right|\leq M. \label{MM}\end{equation}
For $V^j_\Delta(t)$, we have given bases (unitary with respect to \eqref{form}) $u^r_j$, $r=1,\dots,\delta_{j1}k+\delta_{j2}l$, in which multiplication by $\eta$ is represented by the chosen matricial polynomials. Again, we can assume that  over $\Xi_{11}(K)$  and $\Xi_{22}(K)$
\begin{equation} \left|\Phi_{21}(u_1^r)\right|\leq M,\quad \left|\Phi_{12}(u^r_2)\right|\leq M. \label{MMM}\end{equation}
\begin{remark} In both \eqref{MM} and \eqref{MMM}, we can replace $\Phi_{ij}$ with $\Lambda_{ij}$. Given $\delta>0$, we can choose an $M=M(K,\delta)$, such that \eqref{MM} and \eqref{MMM} hold with this $M$ for all $t\in [\delta,2-\delta]$.\label{MMMM}\end{remark}
\par
We now write $\Xi_1(K,R)$ (resp. $\Xi_2(K,R)$) for the subset of $\Sigma_{k,l}(K)$ defined by $c_1(\zeta)=\frac{lR\zeta}{k+l}$ (resp. $c_2(\zeta)=-\frac{kR\zeta}{k+l}$). We define similarly sets $\Xi_{ij}(K,R)$ for $i,j=1,2$. We observe that $\Xi_{ij}(K,R)$ are obtained from $\Xi_{ij}(K)$ by the map $t_{c_j(\zeta)}$ defined in Section \ref{trans}. Let $\sW^1_m$ (resp. $\sW^2_m$) be the tautological bundle over $U_\infty^{[m]}$ (resp. $U_0^{[m]})$. Consider the analogous maps $t_{c_j(\zeta)}$ on $\sW^j_m$, given by \eqref{tc1} or \eqref{tc2} and define  new Hermitian metrics on  $\sW^j_m$ by pulling back the old metric  via $t_{c_j(\zeta)}$. This induces new Hermitian metrics on $\sV^j_{\zeta_0}(t)$, $j=1,2$, over $\sS_{k,l}$. In particular, these are the metrics we shall consider for $(S_1,S_2)\in \Xi_1(K,R)\cap \Xi_2(K,R)$.

We need the following
\begin{lemma} Let $S\in \sC_n$ be defined by $P(\zeta,\eta)=0$ and let $c(\zeta)=z+2x\zeta-\bar{z}\zeta^2$ be its centre. Define the corresponding
{\em centred curve} $S^c$ by $P(\zeta,\eta-c(\zeta))=0$. For any $m\in \oN$ and any $t\in \oC$
there is a 1-1 correspondence between sections of $L^t(m)$ on $S$ and on $S^c$. The correspondence is given by
$$s_0^c(\zeta,\eta)=e^{t(x-\bar{z}\zeta)}s_0(\zeta,\eta-c(\zeta)),\quad s_\infty^c(\zeta,\eta)=e^{t(-x-z/\zeta)}s_\infty(\zeta,\eta-c(\zeta)),$$
 where $s_0,s_\infty$ represent a section of $L^t(m)_{|S}$ in the trivialisation $U_0,U_\infty$.\label{centering}\end{lemma}
 \begin{proof}
  We check that $s_0^c,s_\infty^c$ define a section of $L^t(m)$ on $S^c$:
  \begin{multline*} e^{-t\eta/\zeta}\zeta^{-m}s_0^c(\zeta,\eta)= \zeta^{-m}e^{-tc(\zeta)/\zeta} e^{-t(\eta-c(\zeta))/\zeta}e^{t(x-\bar{z}\zeta)}s_0(\zeta,\eta-c(\zeta))
  =\\ \zeta^{-m}\left(e^{-tc(\zeta)/\zeta}e^{t(x-\bar{z}\zeta)}\right) \left(e^{-t(\eta-c(\zeta))/
\zeta}s_0(\zeta,\eta-c(\zeta))\right)=\\e^{t(-x-z/\zeta)}s_\infty(\zeta,\eta-c(\zeta))=s_\infty^c(\zeta,\eta).
  \end{multline*}
\end{proof}

{\bf Step 3.} We go back to $(S_1,S_2)$ as in Step 1, i.e. $(S_1,S_2) \in \Sigma_{k,l}$ with $R(S_1,S_2)=R$ and $\zeta_{21}=0$. We write $(S^1_1,S_2^1)$ (resp. $(S^2_1,S_2^2)$) for the translation of $S_1$ and $S_2$ by $-c_1(\zeta)$ (resp. $-c_2(\zeta)$). Thus $S_1^1$ and $S_2^2$ have null centres.
Let $u^r_j$ be the basis of $V^j_\Delta(t)$, in which multiplication by $\eta$ is represented by the chosen matricial polynomials. We observe that $u^r_j$ for $S_j$ is obtained, via the formula in Lemma \ref{centering}, from $u^r_j$ for $S_j^j$. Let $v_j^p$, $p=1,\dots,kl$, $j=1,2$, be  unitary bases of $H^0\bigl(S_j^j, L^t(k+l-1)[-D_{\zeta_0}]\bigr)$, with respect to the metrics defined in Step 2. Lemma \ref{centering} gives us bases $\tilde{v}_j^p$ of $H^0\bigl(S_j, L^t(k+l-1)[-D_{\zeta_0}]\bigr)$. With respect to the metrics on $H^0\bigl(S_j, L^t(k+l-1)[-D_{\zeta_0}]\bigr)$, defined just before Lemma \ref{centering}, we have:
\begin{equation} \langle\tilde{v}_1^p,\tilde{v}_1^q\rangle=\delta_{pq}e^{\frac{2lRt}{k+l}}, \quad \langle\tilde{v}_2^p,\tilde{v}_2^q\rangle=\delta_{pq}e^{\frac{2kRt}{k+l}}.\label{deltapq} \end{equation}
\par
For any $u_1^r$ we seek $w_1\in H^0\bigl(S_1, L^t(k+l-1)[-D_{\zeta_0}]\bigr)$ and $w_2\in H^0\bigl(S_1, L^t(k+l-1)[-D_{\zeta_0}]\bigr)$ so that (cf. \eqref{Lambda})
\begin{equation} \Lambda_{11}(w_1)-\Lambda_{12}(w_2)=-\Lambda_{11}(u^r_1),\quad
\Lambda_{21}(w_1)-\Lambda_{22}(w_2)=-\Lambda_{21}(u^r_1),\label{Lambda2}\end{equation}
and similarly for $u_2^r$. We write $w_1=\sum x^p_1\tilde{v}_1^p$ and $w_2=\sum x^p_2\tilde{v}_2^p$ so that \eqref{Lambda2} becomes the matrix equation:
$$\left(\begin{array}{cc}B_{11} & B_{12}\\ B_{21} & B_{22}\end{array}\right)
\left(\begin{array}{c}x_1\\x_2\end{array}\right)=\left(\begin{array}{c}C_1\\C_2\end{array}\right).$$
From \eqref{deltapq}, we know that $B_{11}=e^{\frac{lRt}{k+l}}\cdot I$ and $B_{22}=e^{\frac{kRt}{k+l}}\cdot I$. On the other hand, \eqref{MM}, \eqref{MMM}, Remark \ref{MMMM} and Lemma \ref{centering} imply that all entries of $B_{21}$ are bounded by $Me^{-\frac{lRt}{k+l}}$, while all entries of $B_{12}$  are bounded by $Me^{-\frac{kRt}{k+l}}$. In particular, the matrix $B$ is invertible, if $Rt$ is greater than some $N=N(k,l,M)=N(k,l,K)$. 
This holds for $t\in [\delta,2-\delta]$, if $R$ is sufficiently large. Similarly, if we solve \eqref{Lambda2} with the right-hand side given by $u^r_1$, then $C_1=0$ and every entry of $C_2$ is bounded by $Me^{-\frac{lRt}{k+l}}$. If we solve \eqref{Lambda2} with the right-hand side given by $u^r_2$, then $C_2=0$ and every entry of $C_1$ is bounded by $Me^{-\frac{kRt}{k+l}}$. It follows that, if $t\in [\delta,2-\delta]$ and $R\geq R_0$, then the entries of $x_1$ and $x_2$ satisfy:
\begin{equation}|x^p_1|\leq Me^{-Rt},\quad |x^p_2|\leq Me^{-Rt},\label{estimate_x}\end{equation}
for a new constant $M=M(K,\delta)$.

{\bf Step 4.}
We show that the basis of $H^0\bigl(S_1\cup S_2, L^t(k+l-1)\bigr)$, obtained above, can be replaced by a unitary one. Let $u^1_1,\dots,u_1^k$ and $u^1_2,\dots,u_2^l$ be the (unitary) bases of $H^0\bigl(S_1,L^t(k+l-1)[-\tau(D)]\bigr)$ and $H^0\bigl(S_2,L^t(k+l-1)[-D]\bigr)$, in which the multiplication by $\eta$ gives the chosen matricial polynomials. Step 2 has given us, for $t\in [\delta,2-\delta]$ a basis of $H^0\bigl(S_1\cup S_2, L^t(k+l-1)\bigr)$ of the form 
\begin{equation}(u_1^1+w_1^1,w_2^1),\dots,(u_1^k+w_1^k,w_2^k),(y_1^1,u_2^1+y_2^1),\dots, (y_1^l,u_2^l+y_2^l),\label{base}\end{equation}
where $w_i^r,y_i^s\in H^0\bigl(S_i,L^t(k+l-1)[-D_{\zeta_0}]\bigr)$. We claim that this basis is almost orthonormal with respect to \eqref{form} on $H^0\bigl(S_1\cup S_2, L^t(k+l-1)\bigr)$.
  We use the formula \eqref{c0} for the metric on  $H^0\bigl(S_1\cup S_2, L^t(k+l-1)\bigr)$ and on $H^0\bigl(S_i,L^t(k+l-1)[-\tau^i(D)]\bigr)$, $i=1,2$.  Observe that on $S_1\cup S_2$, this formula can be written as 
\begin{equation} \langle v,w\rangle=\sum_{(\eta,\zeta_1)\in S_1} \Res\frac{v_1\sigma(w_1)(\eta,\zeta_1)}{P(\eta,\zeta_1)} + \sum_{(\eta,\zeta_1)\in S_2} \Res\frac{v_2\sigma(w_2)(\eta,\zeta_1)} { P(\eta,\zeta_1)},
\label{form1}\end{equation}
where $P=P_1P_2$ is the polynomial defining $S=S_1\cup S_2$ and $\zeta_1$ is an arbitrary point of $ \oP^1$. 
\par
Let now $v,w$ be arbitrary sections in  $H^0\bigl(S_1, L^t(k+l-1)\bigr)$. Then $v\sigma(w)$ is a section of $\sO(2k+2l-2)$ on $S_1$, and according to \cite[Lemma(2.16)]{HuMu}, it can be written 
as $\sum_{i=0}^{k+l-1}\eta^if_i(\zeta)$ with $\deg f_i= 2k+2l-2-2i$. This representation is not unique: adding any polynomial of the form $h(\zeta,\eta)P_1(\zeta,\eta)$ defines the same section. 
Nevertheless,
$$\sum_{(\zeta_1,\eta)\in D_{\zeta}}\Res\frac{\bigl(v\sigma(w)\bigr)(\zeta_1,\eta)}{P(\zeta_1,\eta)}$$
does not depend on the representation, as long as $\zeta_1\not \in \pi(\supp S_1\cap S_2)$. With our choice of $\zeta_{21}$, Proposition \ref{good_D} implies that there is an $R_0$ such that, for $R(S_1,S_2)\geq R_0$ and $\bar{B}=\{\zeta;1/2\leq |\zeta|\leq 2\}\cap \pi(\supp S_1\cap S_2)=\emptyset$.  The above discussion is valid for 
$v,w\in H^0\bigl(S_1, L^t(k+l-1)\bigr)$ as well, and, therefore, on the set 
$$\Sigma^0=\{(S_1,S_2)\in \Sigma_{k,l}(K); \zeta_{21}=0, R(S_1,S_2)\geq R_0\},$$
we have well defined quantities 
\begin{equation} N_i(v,w)=\sup_{\zeta\in \overline{B}}\left| \sum_{(\zeta,\eta)\in D_{\zeta}}\Res\frac{\bigl(v\sigma(w)\bigr)(\zeta,\eta)}{P(\zeta,\eta)}\right|,\end{equation}
 for any $v,w\in H^0\bigl(S_i, L^t(k+l-1)\bigr)$, $i=1,2$.
Observe that the $N_i$  equal the corresponding $N_i$ for $v^c, w^c\in H^0\bigl(S_i^c,L^t(n+p-1)\bigr)$,
obtained via Lemma \ref{centering}. The $N_i$ are upper semi-continuous as functions on $\sV^1\oplus\sV^2$ over $\Sigma^0$, and the compactness argument, used in Step 2, guarantees that there is a constant $N=N(k,l,\delta)$ such that 
$$N_i(u_i^r,\tilde{v}_i^p),N_i(\tilde{v}_i^r,\tilde{v}_i^s)\leq N,\quad i=1,2,$$
for all $(S_1,S_2)\in \Sigma_{k,l}(K)$, $t\in [\delta,2-\delta]$, and all $r,p,s$, where the $\tilde{v}_j^p$ are the bases  of $H^0\bigl(S_j,L^t(k+l-1)[-D_{\zeta_0}]\bigr)$, defined in Step 3. Now, the estimate \eqref{estimate_x} shows that the matrix of the form \eqref{form1} evaluated on  the basis \eqref{base} is $N e^{-Rt}$-close to the identity matrix (different $N$). We can, therefore, for any $t\in [\delta,2-\delta]$, use the Gram-Schmidt process and modify the bases $u^1_1,\dots,u_1^k$ of $H^0\bigl(S_1,L^t(k+l-1)[-\tau(D)]\bigr)$ and $u^1_2,\dots,u_2^l$ of $H^0\bigl(S_2,L^t(k+l-1)[-D]\bigr)$
by vectors of length $N e^{-Rt}$ (relative to these bases), so that the solution of \eqref{Lambda2} will be unitary in $H^0\bigl(S_1\cup S_2, L^t(k+l-1)\bigr)$.

{\bf Step 5.}  We prove Theorem \ref{hardest2}, which also proves the second statement of Theorem \ref{hardest}.
We have a unitary basis of $H^0\bigl(S_1\cup S_2, L^t(k+l-1)\bigr)$ of the form \eqref{base}. We rename $u_1^1,\dots,u^k_1,u_2^1,\dots,u^l_2$ as $\hat{\psi}_1,\dots,\hat{\psi}_{k+l}$ and we rename the basis \eqref{base} as ${\psi}_1,\dots,{\psi}_{k+l}$.
The matricial polynomials $\hat{A}(t,\zeta)$ and ${A}(t,\zeta)$ represent multiplication by $\eta$ in the bases $\hat{\psi}_i$ and ${\psi}_i$. The formula \eqref{endom} defines $g(t,\zeta)$ and shows that it is meromorphic in $\zeta$ with only possible singularities at points of $\pi\bigl(\supp S_1\cap S_2\bigr)$. The \eqref{conjugate} shows that, at any point $\zeta\in \oP^1$, such that $\supp D_\zeta$ on $S_1\cup S_2$ consists of $k+l$ distinct points $p_1,\dots,p_k\in S_1$, $p_{k+1},\dots,p_{k+l}\in S_2$ (such points are generic, thanks to Corollary \ref{nomult}), we have
$$  g(t,\zeta)=\bigl[\hat{\psi}^j(p_i)\bigr]^{-1}\bigl[\psi^j(p_i)\bigr].$$
In particular, $g(t,\zeta)$ satisfies the assertion (ii) of Theorem \ref{hardest2}. Moreover, since $\det \bigl[\hat{\psi}^j(p_i)\bigr]$ and $\det \bigl[\psi^j(p_i)\bigr]$ vanish to the same order at any $\pi\bigl(\supp S_1\cap S_2\bigr)$, we conclude that $\det g(t,\zeta)$ is constant and can be assumed to be $1$.
\par
Represent each $u_j^r$ by $(u_j^r)_0$ and $(u_j^r)_\infty$ in $U_0\cap S_j$ and $U_\infty\cap S_j$, $j=1,2$. Let $G$ be a compact subset of $\oP^1-\{\infty\}$ with a nonempty interior. Because of the compactness of $\Xi_{11}(K)$  and $\Xi_{22}(K)$, we have 
\begin{equation} N_j(G)=\sup_r\bigl\{ |(u_j^r)_0(\zeta,\eta)|;\enskip \zeta\in G,\enskip(S_j,\tau^j(D))\in \Xi_{jj}(K)\bigr\}<+\infty.\label{Mj}\end{equation}
Similarly, for every vector $s$ of length one in the restriction of $\sV^1_{\zeta_0}(t)$ to $\Xi_{11}(K)$ or  in the restriction of $\sV^2_{\zeta_0}(t)$ to $\Xi_{22}(K)$, we have
\begin{equation}\sup\bigl\{ |s_0(\zeta,\eta)|;\enskip \zeta\in G \bigr\}\leq O_j(G)\label{NN}\end{equation}
for some finite number $O_j(G)$, $j=1,2$. 
\par
 Consider the sections $u_j^r$ of $H^0\bigl(S_j,L^t(k+l-1)[-\tau^j(D)]\bigr)$ and, as in Step 3, $\tilde{v}_j^p$ of $H^0\bigl(S_j,L^t(k+l-1)[-D_{\zeta_0}]\bigr)$. Let $\tilde{N_j}(G)$,  $\tilde{O}_j(G)$ be the suprema  applied to these sections (for $\zeta\in G$). Lemma \ref{centering} gives:
\begin{equation} \tilde{N}_1(G)\leq N_1(G)e^{-\frac{lRt}{k+l}},\enskip \tilde{N}_2(G)\leq N_2(G)e^{\frac{kRt}{k+l}},\label{tilN}\end{equation}\begin{equation} \tilde{O}_1(G)\leq O_1(G)e^{-\frac{lRt}{k+l}},\enskip \tilde{O}_2(G)\leq O_2(G)e^{\frac{kRt}{k+l}}.\label{tilO} \end{equation}
Now, our basis $\psi^j$ of $H^0\bigl(S_1\cup S_2, L^t(k+l-1)\bigr)$  is of the form \eqref{base}, where  $w_i^r$ and $y_i^s$ are linear combinations of the $\tilde{v}_i^p$ with coefficients  satisfying the estimates \eqref{estimate_x}. Hence
\begin{equation} \sup_{r,s}\bigl\{ \bigl|(w_1^r)_0(\zeta,\eta)\bigr|, \bigl|(y_1^s)_0(\zeta,\eta)\bigr|;\enskip \zeta\in G\bigr\}\leq MO_1(G)e^{-(k+2l) Rt/(k+l)}.\label{suppp}\end{equation}
\begin{equation} \sup_{r,s}\bigl\{\bigl|(w_2^r)_0(\zeta,\eta)\bigr|, \bigl|(y_2^s)_0(\zeta,\eta)\bigr|;\enskip \zeta\in G\bigr\}\leq MO_2(G)e^{-l Rt/(k+l)}.\label{supppp}\end{equation}
Let us write $\psi(\zeta)=\bigl[\psi^j(p_i)\bigr]$ and $\hat{\psi}(\zeta)=\bigl[\hat{\psi}^j(p_i)\bigr]$.
We can also write
$$ \psi(\zeta)=\left(\begin{array}{cc} e^{-l Rt/(k+l)}\cdot 1 & 0\\ 0 & e^{k Rt/(k+l)}\cdot 1\end{array}\right)\left(\begin{array}{cc} C_{11}(\zeta) & C_{12}(\zeta)\\ C_{21}(\zeta) & C_{22}(\zeta)\end{array}\right),$$
where the diagonal blocks have sizes $k\times k$ and $l\times l$. The above estimates imply
$$ \bigl|C_{11}(\zeta)\bigr|,\bigl|C_{22}(\zeta)\bigr|\leq N,\quad \bigl|C_{12}(\zeta)\bigr|,\bigl|C_{21}(\zeta)\bigr|\leq Me^{-\alpha Rt},$$
for all $\zeta\in G$ and all $(S_1,S_2)\in \Sigma_{k,l}(K)$  with $R(S_1,S_2)$ sufficiently large ($N,M,\alpha$ depend only on $k,l,\delta,K,G$). Similarly, we can write
 $$ \hat{\psi}(\zeta)=\left(\begin{array}{cc} e^{-l Rt/(k+l)}\cdot 1 & 0\\ 0 & e^{k Rt/(k+l)}\cdot 1\end{array}\right)\left(\begin{array}{cc} \hat{C}_{11}(\zeta) & 0 \\ 0 & \hat{C}_{22}(\zeta)\end{array}\right),$$
with $\bigl|\hat{C}_{ii}(\zeta)\bigr|$ bounded by $N$, and $\bigl|{C}_{ii}(\zeta)-\hat{C}_{ii}(\zeta)\bigr| \leq Me^{-\alpha Rt}$. Let $C(\zeta)$ and $\hat{C}(\zeta)$ be the matrices with blocks $C_{ij}(\zeta)$ and $\hat{C}_{ij}(\zeta)$ (we omit the $t$-dependence). Then $g(t,\zeta)=\hat{C}(\zeta)^{-1}C(\zeta)$ and since $C(\zeta)$ is uniformly bounded on $G$ and $\det g(t,\zeta)=1$, $\det\hat{C}(\zeta)$ is uniformly  bounded on $G$. Together with the above estimates, this proves the assertion (i) of Theorem \ref{hardest2}.

{\bf Step 6.} We prove the first statement of Theorem \ref{hardest}. We have to show that the Nahm flow corresponding to $L^t(k+l-1)$ on $S_1\cup S_2$ does not have singularities for all $t\in (0,2)$. We know already, from Step 3, that there is an $N=N(k,l,K)$, such that the flow is regular on $(N/R,2)$. Suppose that there is a sequence $(S_1^n,S_2^n)\in
\Sigma_{k,l}(K)$ (with the standing assumption that the total center is zero and $\zeta_{21}=1$) such that the flow corresponding to
$L^t(k+l-2)$ on $S_1^n\cup S_2^n$ has a pole at $\epsilon_n\in (0,N/R_n)$, where $R_n=R(S_1^n, S_2^n)$. Let $P_i^n(\zeta,\eta)=0$ be the equations of
$S_i^n$, $i=1,2$, and consider the rescaled curves $\tilde{S}_i^n$ given by the equations $P_i^n(\zeta,\eta/R_n)=0$. The Nahm flow
on $\tilde{S}_1^n\cup \tilde{S}_2^n$ has a pole at $R_n\epsilon_n\in (0,N)$. On the other hand, we can find a converging subsequence of
$\bigl(\tilde{S}_1^n,\tilde{S}_2^n\bigr)$ and the limit $S^\infty$ is a nilpotent curve or the union of two such curves. In both cases the limit Nahm
flow on $S^\infty$is regular on $(0,+\infty)$. For any spectral curve, the Nahm flow (without the $T_0$-component) corresponding to $L^t(k+l-1)$ is a regular
singular ODE the resonances  of which are determined by the coefficients of the curve. Thus, the usual lower semi-continuity of $\omega_+$, where $[0,\omega_+)$ is the maximal interval of existence of solutions to an ODE, implies that, for curves close enough to  $S^\infty$, the Nahm flow is regular on $(0,N+1)$. This is a contradiction.

\section{The asymptotic region of $\sM_{k,l}$ and Nahm's equations}

We consider now these elements of $\Sigma_{k,l}$ for which the flow $L^t(k+l-1)$ on $S_1\cup S_2$ does not meet the theta divisor for $t\in
(0,2)$.  In other words the corresponding Nahm flow exists for $t\in (0,1]$. According to Theorem \ref{hardest}, this is true in the
asymptotic region of $\Sigma_{k,l}(K)$. Recall, once again, that the flows $T_i(t)$ corresponding to $L^t(k+l-1)$ on $S_1\cup S_2$ have poles at $t=0$ corresponding to the irreducible representation of dimension $k+l$. Let $A(t,\zeta)$ denote the corresponding matricial
polynomials, i.e. $A(t,\zeta)=\bigl(T_2(t)+iT_3(t)\bigr) +2iT_1(t)\zeta + \bigl(T_2(t)-iT_3(t)\bigr)\zeta^2$. Theorem \ref{hardest2}
implies that, as long as $R(S_1,S_2)$ is large enough, there is a meromorphic map $g:\oP^1\rightarrow SL(k+l,\oC)$, with poles at $S_1\cap S_2$, such that
$g(\zeta)A(1,\zeta)g(\zeta)^{-1}=\hat{A}(\zeta)$, where $\hat{A}(\zeta)$ is block-diagonal with the blocks corresponding to line bundles
$L^1(k+l-1)_{|S_1}[-\tau(D)]$ and $L^1(k+l-1)_{|S_2}[-D]$.
\par
We define a space $P$ as the set of pairs  $\bigl(A(t,\zeta),g(\zeta)\bigr)$, where $A(t,\zeta)$, $t\in (0,1]$, is the matricial polynomial
corresponding to the flow $L^t(k+l-1)$ on $S_1\cup S_2$ ($(S_1,S_2)\in \sS_{k,l}$) and $g:\oP^1\rightarrow GL(k+l,\oC)$ is meromorphic with
poles at $S_1\cap S_2$, such that $g(\zeta)A(1,\zeta)g(\zeta)^{-1}=\hat{A}(\zeta)$, where $\hat{A}(\zeta)$ is block-diagonal with the
blocks symmetric, satisfying the reality condition \eqref{+} and corresponding to line bundles $L^1(k+l-1)_{|S_1}[-\tau(D)]$ and $L^1(k+l-1)_{|S_2}[-D]$. The
map $g$ is not unique: the conditions on $\hat{A}(\zeta)$ are preserved by conjugation by block-diagonal matrices $H\in U(k)\times U(l)$
such that the non-central parts of the blocks are orthogonal. Let $M$ be the quotient of $P$ by $O(k)\times O(l)$.
\begin{proposition} There is a canonical embedding of $M$ into $\sM_{k,l}$.\label{M}\end{proposition}
\begin{proof} We already have an embedding on the level of spectral curves. We have to show that an element of $M$ gives also a pair of meromorphic
sections of $L^2$ on $S_1$ and on $S_2$. Let $\bigl(A(t,\zeta),g(\zeta)\bigr)$ represent an element of $M$. Just as at the end of section
\ref{monopole} consider the unique solution $w(t,\zeta)$ of $\frac{d}{dt}w+A_\# w=0$ satisfying $t^{-(k+l-1)/2}w(t,\zeta)\rightarrow
(1,0,\dots,0)^T$ as $t\rightarrow 0$ ($(1,0,\dots,0)^T$ lies in the $-(k+l-1)/2$-eigenspace of the residue of $A_\#$). The vector
$w(\zeta)=w(1,\zeta)$ is cyclic for $A(1,\zeta)$ for any $\zeta$, and similarly $w^T(\zeta)$ is a cyclic covector for $A(1,\zeta)^T$. Hence
$g(\zeta)w(\zeta)$ is cyclic for $\hat{A}(\zeta)$, apart from singularities, and $w^T(\zeta)g^T(\zeta)$ is a cyclic covector for
$\hat{A}(\zeta)^T=\hat{A}(\zeta)$. Therefore the following formula is well-defined on $M$ and associates to
$\bigl(A(t,\zeta),g(\zeta)\bigr)$ a meromorphic function on $(S_1\cup S_2)-\pi^{-1}(\infty)$:
\begin{eqnarray}\nu_0(\zeta,\eta)& = & w(\zeta)^Tg^T(\zeta)g(\zeta) \bigl(\eta-A(1,\zeta)\bigr)_{\text{adj}}w(\zeta)=  \label{singnumerator}\\
 & & w(\zeta)^Tg^T(\zeta)\bigl(\eta-\hat{A}(\zeta)\bigr)_{\text{adj}}g(\zeta)w(\zeta). \nonumber\end{eqnarray}
Arguments such as in \cite{Hu0} show that this defines a (meromorphic) section of $L^2$ on $S_1\cup S_2$ and Theorem \ref{hardest2}
shows that $\nu_0$ restricted to $S_1$ and to $S_2$ have correct divisors, i.e. $D-\tau(D)$ on $S_1$ and $\tau(D)-D$ on $S_2$.
\par
Finally, it is clear that $g(\zeta)$ and $Hg(\zeta)$, where $H$ is block-diagonal with each block central, give different $\nu_0$ unless
$H$ is orthogonal. Therefore the map is an embedding.
\end{proof}

From the proof we obtain an interpretation of the biholomorphism $\sM_{k,l}^{\zeta_0}\simeq
\text{Rat}_k\bigl(\oP^1\bigr)\times\text{Rat}_l\bigl(\oP^1\bigr)$ of Theorem \ref{existence} in terms of Nahm's equations:
\begin{corollary} The composition of the embedding $M\hookrightarrow \sM_{k,l}$ with the biholomorphism $\sM_{k,l}^{\zeta_0}\simeq
\text{Rat}_k\bigl(\oP^1\bigr)\times\text{Rat}_l\bigl(\oP^1\bigr)$ is given by $\bigl(A(t,\zeta),g(\zeta)\bigr) \rightarrow
\left(\frac{p_1(z)}{q_1(z)}, \frac{p_2(z)}{q_2(z)}\right)$, where $q_1,q_2$ are the equations of $S_1,S_2$ at $\zeta=\zeta_0$ and $p_1,p_2$
are defined by
$$ p_1(z)\equiv \nu_0(\zeta_0,z)\mod q_1(z),\quad\quad p_2(z)\equiv \nu_0(\zeta_0,z)\mod q_2(z),$$
with $\nu_0$ given by \eqref{singnumerator}.\hfill $\Box$\end{corollary}
For every $\zeta_0\in \oP^1$ we now define a map from a subset of $M$ (i.e. from a subset of $\sM_{k,l}$) to the monopole moduli space
$\sM_{k+l}$. This map is simply given by a corresponding map on the rational functions. Let $\left(\frac{p_1(z)}{q_1(z)},
\frac{p_2(z)}{q_2(z)}\right)\in \text{Rat}_k\bigl(\oP^1\bigr)\times\text{Rat}_l\bigl(\oP^1\bigr)$ and assume that $q_1$ and $q_2$ are
relatively prime. We define a rational map $\frac{P(z)}{Q(z)}$ of degree $k+l$ by $Q(z)=q_1(z)q_2(z)$ and $P(z)$ as the unique polynomial
of degree $k+l-1$ such that $P(z)\equiv p_1(z)\mod q_1(z)$ and $P(z)\equiv p_2(z)\mod q_2(z)$. The map
$$\left(\frac{p_1(z)}{q_1(z)}, \frac{p_2(z)}{q_2(z)}\right)\longmapsto \frac{P(z)}{Q(z)}$$
induces a map from the corresponding region of $\sM_{k,l}$ to $\sM_{k+l}$. We shall abuse the notation and write
$$ \Phi_{\zeta_0}:\sM_{k,l}\longrightarrow \sM_{k+l}$$
for this map (even that it is not defined on all of $\sM_{k,l}$). It is clearly holomorphic for the chosen complex structure and preserves
the corresponding complex symplectic form. We also observe that generically $\Phi_{\zeta_0}$ is
$\left(\begin{array}{c}k+l\\k\end{array}\right)$  to  $1$.
\par
The region on which $\Phi_{\zeta_0}$ is defined contains an open dense subset of $M$ (given by the condition $\zeta_0\not\in \pi(S_1\cap
S_2)$) and we we wish to give a description of $\Phi_{\zeta_0}$ in terms of solutions to Nahm's equations. First of all, the map which
associates to an $ \bigl[A(t,\zeta),g(\zeta)\bigr)]\in M$ the rational function $\frac{P(z)}{Q(z)}$ is given,  by the  discussion above, by
\begin{equation} \bigl(A(t,\zeta),g(\zeta)\bigr)\longmapsto
w(\zeta_0)^Tg^T(\zeta_0)\bigl(z-\hat{A}(\zeta_0)\bigr)^{-1}g(\zeta_0)w(\zeta_0),\label{map}\end{equation} where $w(\zeta)$ is defined as in
the proof of Proposition \ref{M}.
\par
To obtain a solution to Nahm's equations, corresponding to $P(z)/Q(z)$, directly from $\bigl[\bigl(A(t,\zeta),g(\zeta)\bigr)]\in M$ we
proceed as follows. Thanks to the $SO(3)$-action, we can assume, without loss of generality, that $\zeta_0=0$. We then split the Nahm
equations into a complex one and a real one, as in \eqref{complex} and \eqref{real}. Then $\beta(t)=A(t,\zeta_0)$ and
$\alpha(t)=A_\#(t,\zeta_0)$. Since $\zeta_0\not\in \pi(S_1\cap S_2)$, $g(\zeta_0)$ is a regular matrix which conjugates $\beta(1)$ to a
symmetric and block-diagonal matrix $B$. Extend $g(\zeta_0)$ to a smooth path $g(t)\in Gl(n,\oC)$, for $t\in [0,1]$, with $g(t)=1$ for
$t\leq 1/2$, $g(1)=g(\zeta_0)$ and $\tilde{\alpha}(t)=g(t)\alpha(t) g(t)^{-1}-\dot{g}(t)g(t)^{-1}$ being symmetric at $t=1$. Let
$\tilde{\beta}(t)=g(t)\beta(t)g(t)^{-1}$ and extend $\tilde{\alpha},\tilde{\beta}$ to $[0,2]$ by symmetry. We obtain a smooth solution to
the complex Nahm equation on $[0,2]$ with boundary conditions of an element of $\sM_{k+l}$. We can now find, as in \cite{Don}, a unique
solution to the real equation via a complex gauge transformation $G(t)$ which is identity at $t=0,2$. The resulting solution is the value of
$\Phi_{\zeta_0}$ at $ \bigl[\bigl(A(t,\zeta),g(\zeta)\bigr)]$.

\medskip

We are now going to show that asymptotically the map $\Phi_{\zeta_0}$ is  exponentially close to the identity. For this we need to restrict
the asymptotic region and define it directly in terms of rational functions, as in Corollary \ref{eest}.
\begin{definition} Let  $\zeta_0\in \oP^1$ and $K>0$. We denote by $\sM_{k,l}^{\zeta_0}(K)$ the subset of $\sM_{k,l}^{\zeta_0}$ corresponding to $\left(\frac{p_1(z)}{q_1(z)},\frac{p_2(z)}{q_2(z)}\right)\in {\rm Rat}_k\bigl(\oP^1\bigr)\times {\rm Rat}_l\bigl(\oP^1\bigr)$ which satisfy:
\begin{itemize}
\item[(i)] Any zero of $q_1(z)$ is at least distance $1$ apart from any zero of $q_2(z)$.
\item[(ii)] Any two zeros of $q_1(z)$ (resp. of $q_2(z)$) are distance at most $2K$ apart.
\item[(iii)] If $\beta_1,\beta_2$ are two zeros of $q_1(z)$ (resp. of $q_2(z)$), then $\bigl|\ln |p_1(\beta_1)|-\ln |p_1(\beta_2)|\bigr|\leq 2K$ (resp. $\bigl|\ln |p_2(\beta_1)|-\ln |p_2(\beta_2)|\bigr|\leq 2K$).
\end{itemize}\label{last_asym}\end{definition}

In other words, $\sM_{k,l}^{\zeta_0}(K)$ corresponds to pairs of rational functions, which are within fixed "distance" from $\left(\frac{e^{a_1}}{(z-b_1)^k},\frac{e^{a_2}}{(z-b_2)^l}\right)$, where  $b_1=\sum \beta_i^1/k$, $b_2=\sum \beta_i^2/l$, $a_1=\sum \ln|p_1(\beta_i^1)| /k$, $a_2=\sum \ln|p_2(\beta_i^2)| /l$ (here where $\beta_1^1,\dots,\beta_k^1$ (resp.  $\beta_1^2,\dots,\beta_l^2$) are the roots of $q_1(z)$ (resp. $q_2(z)$)). For an $m\in\sM_{k,l}^{\zeta_0}$, let us define $$R^{\zeta_0}(m)=\min \{ \bigl|\beta_i^1-\beta_j^2\bigr|;\enskip i=1,\dots,k, \enskip j=1,\dots, l\}.$$ If $m=(S_1,\nu_1,S_2,\nu_2)$, then we obviously have $R(S_1,S_2)\geq R^{\zeta_0}(m)$. With these preliminaries, we have:
\begin{theorem} For every $K>0$, there exist positive constants $R_0,\alpha,C$ such that the map $\Phi_{\zeta_0}$ satisfies the following estimates in the
region of $\sM_{k,l}^{\zeta_0}(K)$, where $R^{\zeta_0}(m)\geq R_0$ and $\zeta_0$ is at least distance $1/2$ from the roots of $(b_1-b_2)+2(a_1-a_2)\zeta- (\bar{b}_1-\bar{b}_2)\zeta^2$.
 Let $\Phi_{\zeta_0}(S_1,\nu_1,S_2,\nu_2)=(S,\nu)$. Then $d(S,S_1\cup
S_2)\leq Ce^{-\alpha R}$. Moreover, the numerators $\tilde{p}_\zeta(z), p_\zeta(z)$ of the rational functions of degree $k+l$, corresponding to
$(S_1,\nu_1,S_2,\nu_2)$ and to $(S,\nu)$ and direction $\zeta$ (so that $p_{\zeta_0}(z)=\tilde{p}_{\zeta_0}(z)$), are also exponentially close
for  $\zeta$ sufficiently close to $\zeta_0$ in the sense that $\bigl|p_\zeta(\beta_i)-\hat{p}_\zeta(\hat{\beta}_i)\bigr|\leq Ce^{-\alpha
R}|\tilde{p}_\zeta(\hat{\beta}_i)\bigr|$, where $\hat{\beta}_i,\beta_i$, $i=1,\dots,k+l$, are the $\eta$-coordinates of points of $S_1\cup
S_2$ and of $S$ lying above $\zeta$.\label{compare1}
\end{theorem}
\begin{proof} According to Theorems \ref{hardest} and \ref{hardest2} (and Remark \ref{good_rat}), in the region under consideration, we can conjugate the flow $A(t,\zeta)$ by a unitary $u(t)$, $u(0)=1$, so
that $A(1,\zeta)$ is $Ce^{-\alpha R}$-close to a block-diagonal and symmetric $\hat{A}(\zeta)$ (and satisfying the reality condition
\eqref{+}). Moreover, in the notation of Theorem \ref{hardest2}, the matrix $g(\zeta)$ which conjugates $A(1,\zeta)$ to
$\hat{A}(\zeta)$ is, for $\zeta$ close to
$\zeta_0$,  $Ce^{-\alpha R}$-close to identity.  The solutions $\tilde{\alpha},\tilde{\beta}$, defined before Definition \ref{last_asym}, to the complex
Nahm equation on $[0,2]$ are then exponentially close to satisfying the real equation, in the sense that the difference of the two sides in
\eqref{real} is bounded  by $Ce^{-\alpha R}$. It follows then, using Lemma 2.10 in \cite{Don}, that the complex gauge transformation
$G(t)$, $G(0)=G(2)=1$, which solves the real equation is $Ce^{-\alpha R}$-close to a unitary gauge transformation, uniformly on $[0,2]$,
and $\dot{G}G^{-1}$ is uniformly  $Ce^{-\alpha R}$-close to a skew-hermitian matrix. The result follows.
\end{proof}

\section{Comparison of metrics}

We wish to show that the (local) biholomorphism $\Phi_{\zeta_0}$ of the previous section is very close to being an isometry when the
clusters are far apart. Recall the definition \ref{last_asym} of the region $\sM_{k,l}^{\zeta_0}(K)$, and the notation following that definition. Then:
\begin{theorem} Let $g$ and $\tilde{g}$ be the hyperk\"ahler metrics on $\sM_{k+l}$ and $\sM_{k,l}$, respectively.  For every $K>0$, there exist positive constants $R_0,\alpha,C$ such that,  in the
region of $\sM_{k,l}^{\zeta_0}(K)$, where $R^{\zeta_0}(m)\geq R_0$ and $\zeta_0$ is at least distance $1/2$ from the roots of $(b_1-b_2)+2(a_1-a_2)\zeta- (\bar{b}_1-\bar{b}_2)\zeta^2$, the following estimate holds:
$$ \bigl\|\Phi_{\zeta_0}^\ast g-\tilde{g}\bigr\|\leq Ce^{-\alpha R}.$$
\label{compare2}\end{theorem}

The remainder of the section is devoted to the proof of this theorem.
\par
The metric \eqref{metric} on $\sM_{k+l}$ is given in terms of solutions to infinitesimal Nahm's equations \eqref{tangent}.  
Things are more complicated for $\sM_{k,l}$. Although, we have a description of $\sM_{k,l}$ as a {\em space} of solutions to Nahm's equations, it is not a {\em moduli space} (i.e. there is no gauge group involved). In particular, in our description of $\sM_{k,l}$, a tangent vector is a triple $(\tilde{t}_1,\tilde{t}_2,\tilde{t}_3)$ on $[0,1]$ satisfying only the last three equations in \eqref{tangent}, with $\tilde{t}_0=0$ (and, of course satisfying additional restrictions, since we allow to vary spectral curves only in special directions). Nevertheless, the first equation in \eqref{tangent} arises only by adding an infinitesimal gauge transformation, and this has no effect on the K\"ahler form corresponding to $\zeta_0$. 
This fact can be interpreted by trivialising the twistor space of any moduli space of solutions to Nahm's equations, such as $\sM_{k+l}$. For a solution $\bigl(T_1(t),T_2(t),T_3(t)\bigr)$, set $A(t,\zeta)=\bigl(T_2(t)+ iT_3(t)\bigr)+iT_1(t)\zeta+\bigl(T_2(t)- iT_3(t)\bigr)\zeta^2$, $A_\#(t,\zeta)=iT_1(t)+\bigl(T_2(t)- iT_3(t)\bigr)\zeta$ for $\zeta\neq \infty$, and $\tilde{A}(t,\zeta)=\bigl(T_2(t)+ iT_3(t)\bigr)/\zeta^2+iT_1(t)/\zeta+\bigl(T_2(t)- iT_3(t)\bigr)$, $\tilde{A}_\#(t,\zeta)=-iT_1(t)+\bigl(T_2(t)+iT_3(t)\bigr)/\zeta$ for $\zeta\neq 0$. Then, over $\zeta\neq 0,\infty$, we have $\tilde{A}=A/\zeta^2$, $\tilde{A}_\#=A_\#-A/\zeta$. The fibrewise complex symplectic form, given by  \eqref{sympl_form}, on the twistor spaces $Z\bigl(\sM_{k+l}\bigr)$ of  $\sM_{k+l}$ is then equal to
\begin{equation}\Omega_\zeta= \int_0^2 dA_\#(t,\zeta)\wedge dA(t,\zeta).\label{form_last}\end{equation}
The K\"ahler form $\omega_1$,  corresponding to the complex structure $I_{0}$, is then the linear term in the expansion of  $\Omega_\zeta$ in $\zeta$.
\par
We can give a similar interpretation of the complex symplectic form $\tilde{\Omega}_\zeta$ and the K\"ahler form $\tilde{\omega}_1$ on $\sM_{k,l}$. From the previous section, a solution  $\bigl(T_1(t),T_2(t),T_3(t)\bigr)$ to Nahm's equations on $(0,1]$, corresponding to a point of $\sM_{k,l}$, defines a meromorphic section of $L^2$ on $S_1\cup S_2$ by first combining the matrices $T_i$ into a matricial polynomial $A(t,\zeta)$, as above, and then conjugating $A(1,\zeta)$ by a meromorphic $g(\zeta)$. If we extend, for $\zeta$ close to $0$,  $g(\zeta)$ to a path $g(\cdot,\zeta):[0,1]\rightarrow SL(n,\cx)$ and define $A_\#(t,\zeta)$ as for $\sM_{k+l}$, then the form $\tilde{\Omega}_\zeta$ is equal to
\begin{equation} 2\int_0^1 d\left( g(t,\zeta)A_\#(t,\zeta)g(t,\zeta)^{-1}-\frac{dg(t,\zeta)}{dt}g(t,\zeta)^{-1}\right)\wedge d\left( g(t,\zeta)A(t,\zeta)g(t,\zeta)^{-1}\right).\label{lform}\end{equation}
Again, $\tilde{\omega}_1$ is the $\zeta$-coefficient of this expression. To estimate $d\Phi_{\zeta_0}$, we use the $SO(3)$-action to assume that $\zeta_0=0$. We observe, directly from definitions $\Phi_{0}$ is not only a biholomorphism, but that
it also respects the complex symplectic forms $\Omega_0$ and $\tilde{\Omega}_0$. Thus, to prove the theorem, it suffices to show that $\Phi_0^\ast\Omega_\zeta$, evaluated on vectors of length $1$ in $\tilde{g}$, is exponentially close to $\tilde{\Omega}_\zeta$ for $\zeta$ close to $0$. Equivalently, we can evaluate on tangent vectors $v$, such that $d\Phi_0(v)$ has length $1$ in the metric $g$.  
\par
Furthermore, the above expressions of the forms ${\Omega}_\zeta$ and $\tilde{\Omega}_\zeta$ do not depend on adding an infinitesimal gauge transformation (equal to zero at both ends of the interval) to a tangent vector. This means, in practice, that it does not matter, whether we consider tangent vectors as being quadruples $(t_0,t_1,t_2,t_3)$ satisfying \eqref{tangent}, or triples $(t_1,t_2,t_3)$ satisfying only the last three equations in \eqref{tangent} (with $t_0=0$).
\par
We now consider  a unit tangent vector to $N_{k,l}$, i.e.  solutions $(\check{t}_0,\check{t}_1,\check{t}_2,\check{t}_3)$ to equations \eqref{tangent} on $[-1,0]\cup [0,1]$. The asymptotic region under consideration corresponds to an asymptotic region of  $N_{k,l}$, and there we have $C^0$-bounds on  tangent vectors, obtained as in \cite[pp. 316--318]{CMP2}. From a tangent vector to $N_{k,l}$, we obtain a tangent vector to $\sM_{k,l}$, as an infinitesimal solution $(\tilde{t}_0,\tilde{t}_1,\tilde{t}_2,\tilde{t}_3)$ to Nahm's equations on $[0,1]$. This is done as an infinitesimal version of the proof of Theorems \ref{hardest} and \ref{hardest2} (this is straightforward but rather long and we shall leave out the details), and the estimates, applied to the unit tangent bundle of the compact sets considered there, show that: (i) there is a pointwise $C^0$-bound on the $t_i$, (ii) $\tilde{t}_i(1)$ are exponentially close to being symmetric, and (iii) the infinitesimal variations of $g(t,\zeta)$ and $\frac{dg(t,\zeta)}{dt}$ are exponentially small for $\zeta$ close to $0$. Furthermore, the following expression (which has nothing to do with the metric $\tilde{g}$)
\begin{equation}N(\tilde{t})=-2\sum_{i=0}^3\int_0^1 \tr \tilde{t}_i^2\end{equation}
is $O(1/R)$ close to $1$ (essentially, by integrating the $O(e^{-\alpha Rs})$-difference between $\tilde{t}_i(s)$ and $\check{t}_i(s)$).
\par
Now, an infinitesimal version of the proof of Theorem \ref{compare1} (Lemma 2.10 in \cite{BielAGAG} is now replaced by arguments on p. 152 in \cite{Biel1}) produces a tangent vector $(t_0,t_1,t_2,t_3)$ to $\sM_{k+l}$, which is pointwise exponentially close to $(\tilde{t}_0,\tilde{t}_1,\tilde{t}_2,\tilde{t}_3)$.
The estimate on $N(\tilde{t})$, together with a pointwise bound on $\tilde{t}_i$, shows that the length of $(t_0,t_1,t_2,t_3)$ in the metric $g$ is $O(1/R)$ close to $1$. Hence, if we reverse the steps and assume that the $(t_0,t_1,t_2,t_3)$ thus obtained has length $1$, then $(\tilde{t}_0,\tilde{t}_1,\tilde{t}_2,\tilde{t}_3)$ is still  exponentially close to $(t_0,t_1,t_2,t_3)$ and the pointwise bound on $\tilde{t}_i(s)$ and exponential bound on the corresponding infinitesimal variations of $g(t,\zeta)$ and $\frac{dg(t,\zeta)}{dt}$ remain valid. This, together with the estimates on $A(t,\zeta)$ in the proof of Theorems \ref{hardest} and \ref{hardest2}, shows that \eqref{lform} evaluated on two vectors  $(\tilde{t}_0,\tilde{t}_1,\tilde{t}_2,\tilde{t}_3)$ is exponentially close to \eqref{form_last} evaluated on two unit vectors $(t_0,t_1,t_2,t_3)$. This completes the proof.

\begin{remark} The spaces $N_{k,l}$ and $\sM_{k,l}$ are also biholomorphic for a fixed complex structure $I_{\zeta_0}$ (see the definition of the map $T$ in Section \ref{metricsection}).  The above proof shows that, in the asymptotic region of Theorem \ref{compare2}, this biholomorphism is $O(1/R)$-close to being an isometry. This is again (cf. remark \ref{TNUT}) analogous to the behaviour of the Taub-NUT metrics with positive and with negative mass parameter.\label{N-M}\end{remark}

\section{Concluding remarks}

{\bf 13.1}  It would be interesting to derive the hyperk\"ahler metric on $\sM_{k,l}$ from physical principles, i.e. as a Lagrangian on
pairs of monopoles of charges $k$ and $l$ with a relative electric charge.

\medskip

{\bf 13.2} The metric on $\sM_{k,l}$ can be constructed via the {\em generalised Legendre transform} of Lindstr\"om and Ro\v{c}ek \cite{LR1,LR2}, analogously to the monopole metric \cite{IR,Hough,HMR}. This, and further twistor constructions, will be discussed elsewhere.

\medskip

{\bf 13.3} The constraints on spectral curves in $\Sigma_{k,l}$ are those for $SU(2)$-calorons of charge $(k,l)$ \cite{NyS,ChH}. Is there any physics behind this?

\medskip

{\bf 13.4} As mentioned in the introduction we could not give a description of $\sM_{k,l}$ as a moduli space of Nahm's equations.
Nevertheless there is an analogy with the description of the Gibbons-Manton metric in \cite{CMP1}. For $(S_1,S_2)\in \Sigma_{k,l}$ we would
like to consider the flow $L^s(k+l-2)$ on $S_1\cup S_2$ for all $s\geq 0$. The (unique) compactification (as the moduli space of
semi-stable admissible sheaves) of $J^{g-1}(S_1\cup S_2)$ has a stratum (of smallest dimension) isomorphic to $J^{g_1-1}(S_1)\times
J^{g_2-1}(S_2)$. From the proof of Theorem \ref{hardest} we know that the flow $L^s(k+l-2)$ approaches the flow $L^s(k+l-2)[-\tau(D)]\oplus
L^s(k+l-2)[-D]$ on this boundary stratum as $s\rightarrow +\infty$. Can one obtain $\sM_{k,l}$ as a moduli space of solutions to Nahm's
equations on $[0,+\infty)$ with the corresponding behaviour as $s\rightarrow +\infty$? The Nahm flow will have singularities, so this is
certainly not obvious.

\medskip

{\bf 13.5}  We defined, for every complex structure, a (finite-to-one) biholomorphism $\Phi_\zeta$ between open domains of $\sM_{k,l}$ and
of $\sM_{k+l}$. On the other hand, we have, also for every complex structure, a biholomorphism $\Psi_\zeta$ between an open domain of
$\sM_{k,l}$ and $\sM_k\times \sM_l$, namely the identity on pairs of rational functions. Given Proposition \ref{first} or the arguments in
the proof of Proposition \ref{convergence} and Remark \ref{N-M}, we expect also $\Psi_\zeta$ to be an asymptotic isometry. To obtain a precise rate of
approximation requires a more precise analysis of convergence in Proposition \ref{convergence}, but we expect, by analogy with the
Gibbons-Manton metric, that the metrics on $\sM_{k,l}$ and on $\sM_k\times \sM_l$ are $O(1/R)$-close.

\medskip

{\bf 13.6} Finally, let us address the question of more than two clusters. As mentioned in Introduction, it is clear how to define the
``moduli space" $\sM_{n_1,\dots,n_s}$ of $s$ clusters with magnetic charges $n_1,\dots n_s$, $n_1+\dots +n_s=n$. We need $s$ spectral curves
$S_i\in |\sO(2n_i|$ with $S_i\cap S_j=D_{ij}\cup D_{ji}$, $D_{ji}=\tau(D_{ij})$, and $s$ sections $\nu_i$ of $L^2\bigl[\sum_{j\neq
i}(D_{ji}-D_{ij})\bigr]$ on every $S_i$. They need to satisfy conditions analogous to those for $\sM_{k,l}$.
\par
We also can define a pseudo-hyperk\"ahler metric on $\sM_{n_1,\dots,n_s}$ just as for $\sM_{k,l}$ and even to argue that a map
$\Phi_{\zeta}$ to $\sM_n$ is a biholomorphism. One needs to show that the images of maps $\Phi_{\zeta}$ for different $\zeta$ cover the asymptotic
region of $\sM_n$, i.e. to prove an analogue of Theorem \ref{existence} for $s$ clusters, and this might be hard, since we do not know what the analogue of $N_{k,l}$ should be. Nevertheless, to prove that $\Phi_{\zeta}$ is exponentially close to being an isometry in the asymptotic region of
$\sM_{n_1,\dots,n_s}$ one does not need to rely on the arguments given here. In principle, one could try (also for the case of two clusters) to do everything in terms of theta functions of the spectral curves.

\begin{ack} A Humboldt Fellowship, during which a part of this work has been carried out, is  gratefully acknowledged.\end{ack}

\end{document}